\documentclass[final,10pt]{IEEEtran}
\usepackage[dvips,all,color]{xy}
\usepackage{subfig}
\usepackage[final]{graphicx}
\usepackage{amsmath,amssymb}
\usepackage[dvips,svgnames,x11names]{xcolor}
\usepackage{dsfont}
\usepackage[linesnumbered,ruled,vlined,longend]{algorithm2e}
\usepackage{mathrsfs}
\usepackage{times}
\usepackage{stmaryrd}
\usepackage{psfrag}
\usepackage{shading}
\usepackage{flushend}
\newcommand{\aux}{\mathds{A}}

\newtheorem{cor}{Corollary}
\newtheorem{lem}{Lemma}
\newtheorem{prop}{Proposition}
\newtheorem{defn}{Definition}
\newtheorem{example}{Example}
\newtheorem{rem}{Remark}
\newtheorem{notn}{Notation}
\newcommand{\HA}{\boldsymbol{\mathds{H}}}
\newcommand{\Hi}{\mathds{H}_{-1}}
\newcommand{\W}{\mathbf{i}_\circ}
\newcommand{\sync}{ \varotimes }
\newcommand{\pfsum}{ \textbf{+} }

\renewcommand{\and}{\textrm{\sffamily\textsc{and}}}
\newcommand{\pti}{\mathds{G}}
\newcommand{\etal}{\textit{et} \mspace{3mu} \textit{al.}}
\newcommand{\Q}{\mathscr{P}}
\newcommand{\Mblue}{\color{MidnightBlue}}
\newcommand{\Nblue}{\color{DodgerBlue}}
\newcommand{\Dgreen}{\color{DarkGreen}}
\newcommand{\BRed}{\color{DarkRed}}

\newcommand{\Dblue}{\color{NavyBlue}}
\newcommand{\Darkred}{\color{DarkRed}}
\newcommand{\Aqua}{\color{DarkCyan}}
\title{
\colorbox{AliceBlue}{\bf \sffamily \itshape \footnotesize Submitted For Review For Possible Publication Elsewhere: Journal Reference To Be Added When Available}\\
Pattern Classification In Symbolic Streams\\via Semantic Annihilation of Information}
\author{ 
Ishanu~Chattopadhyay~\IEEEauthorrefmark{1}, Yicheng~Wen~\IEEEauthorrefmark{2} and Asok~Ray~\IEEEauthorrefmark{3}
\IEEEcompsocitemizethanks{
\IEEEcompsocthanksitem\IEEEauthorrefmark{1} Corresponding Author, email: ixc128@psu.edu
\IEEEcompsocthanksitem\IEEEauthorrefmark{2} email: yxw167@psu.edu
\IEEEcompsocthanksitem\IEEEauthorrefmark{3} email: axr2@psu.edu
\IEEEcompsocthanksitem Authors are with the department of Mechanical Engineering, The Pennsylvania State University, University Park, PA 16802 , USA
\IEEEcompsocthanksitem This work has been supported in part by the U.S. Army Research Laboratory and the U.S. Army Research Office under Grant No. W911NF-07-1-0376 and by the Office of Naval Research under Grant No. N00014-09-1-0688. Any opinions, findings and conclusions or recommendations expressed in this publication are those of the authors and do not necessarily reflect the views of the sponsoring agencies.} 
} 
\begin{document}
\maketitle
\begin{abstract}                          
We propose a technique for  pattern  classification  in symbolic streams via selective erasure of observed symbols, in cases where the patterns of interest are  represented as Probabilistic Finite State Automata (PFSA). We define an additive abelian group for  a slightly restricted subset of probabilistic finite state automata (PFSA),  and the group sum  is
  used to formulate pattern-specific  semantic annihilators. The annihilators attempt to identify pre-specified patterns via  removal of essentially all inter-symbol correlations from observed sequences, thereby  turning them into symbolic white noise. Thus a perfect annihilation corresponds to a perfect pattern match. This approach of classification via information annihilation is shown to be strictly advantageous, with theoretical guarantees, for a large class of  PFSA models. The results are supported by simulation experiments.
\end{abstract}
%
\allowdisplaybreaks{
%
%
\section{Introduction and Motivation}
The principal focus of this work is the development of an efficient algorithm for identifying pre-specified patterns of interest
in observed symbolic data streams, where the patterns are represented as Probabilistic Finite State Automata (PFSA) over pre-defined symbolic alphabets.

A finite state automaton (FSA) is essentially a finite graph where the nodes are known as states and the edges are known as transitions, which
are labeled with letters from an alphabet.
A string or a symbol string generated by a FSA is a sequence of symbols belonging to an alphabet, which are generated by stepping through a series of transitions in the graph.
Probabilistic finite state automata, considered in this paper, are finite state machines with probabilities associated  with the transitions.
PFSA have extensively studied as an efficient framework for learning the causal structure of observed dynamical behavior~\cite{Mu96}. This is an example of
inductive inference~\cite{AS83}, defined   as the \textit{process of hypothesizing a general rule from examples}.
In this paper, we are concerned with the special case, where the \textit{inferred general rule} takes the form of a PFSA, and the examples are drawn from a
stochastic regular language. Conceptually, in such scenarios, one is
trying to learn the \textit{structure inside of some \textit{black box}, which is continuously emitting symbols}~\cite{Mu96}.
The system of interest may emit a continuous valued signal; which must be then adequately partitioned to yield a symbolic stream.
Note that such partitioning is merely \textit{quantization} and not \textit{data-labeling},
and several approaches for efficient symbolization have been reported~\cite{RR06}.

Probabilistic automata are more general compared to their non-probabilistic counterparts~\cite{LR05}, and are more suited to modeling stochastic dynamics.
It is important to distinguish between the PFSA models considered in this paper, and the ones considered by Paz~\cite{P71}, and in the detailed recent review by Vidal $\etal$~\cite{VTCC05}.
In the latter framework, symbol generation probabilities are not specified, and we have a distribution over the possible end states, for a given  initial state and an observed symbol. In
the models considered in this paper, symbol generation is probabilistic, but the end state for a given initial state, and a generated symbol is unique.
Unfortunately, authors have referred to both these formalisms as \textit{probabilistic finite state automata} in the literature. The work presented here specifically considers
the latter modeling paradigm considered and formalized in \cite{Mu96,SSC02,SS04,CR08}.

The case for using PFSA as pattern classification tools is compelling.
Finite automata are simple, 
%
%
 and the sample and time complexity required for learning them can be easily characterized. This yields significant  computational advantage
 in time constrained applications, over more expressive frameworks such as belief (Bayesian) networks~\cite{P88,HGC94} or
 Probabilistic Context Free Grammars (PCFG)~\cite{JLM92,VTCC205} (also see \cite{CS07} for a general approach to identifying PCFGs from observations) and hidden Markov models (HMMs)~\cite{R89}. Also, from a
computational viewpoint, it is possible to
come up with provably efficient algorithms to optimally learn PFSA, whereas ``optimally learning
HMMs is often  hard''~\cite{Mu96}.
Furthermore, most reported work on HMMs~\cite{R89,KHM96, BJEG97} assumes the model structure or topology is specified in advance, and the learning procedure is merely training, $i.e.$,  finding the right transition probabilities.
For PFSA based analysis, researchers have investigated the more general problem of learning the model topology, as well as the transition probabilities, which implies  that such  analysis can then be applied to domains where there is   no prior knowledge as to what the correct structure might look like~\cite{CY89}.
%

\begin{figure}[t]
\centering
\psfrag{P}[lc]{ \txt{\Dgreen Physical \\\Dgreen Process $G$\\$\phantom{.}$}}
\psfrag{S}[cc]{\txt{ Simpler \\  Compression}}
\psfrag{C}[cc]{\txt{ Faster\\  Convergence}}
\psfrag{H}[cc]{\Mblue \txt{ H}}
\psfrag{D}[lc]{$\phantom{X}$\txt{\Dblue  Sensed \\ \Dblue Symbolic Stream\\ $\phantom{H}$}}
\psfrag{W}[cc]{\bf \Darkred \scriptsize  \txt{Symbolic \\White Noise}}
\psfrag{G}[tl]{ \Mblue \txt{$\phantom{.}$\\\Dblue Stream Generated by annihilator\\\Dblue  $H \ \mathrm{where} \ G \pfsum H = \textbf{0} $ \itshape (additive  zero)}}
 \includegraphics[width=3.25in]{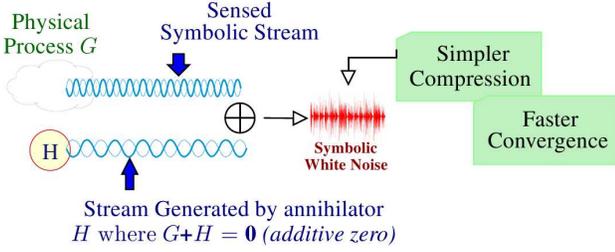}
\caption{Concept of information annihilation: \textit{Addition} of symbolic streams to yield \textit{symbolic white noise} }\label{figconcept1}\vspace{0pt}
\end{figure}

Although the reported PFSA construction algorithms \cite{SS04,SSC02,Mu96} (\textit{referred to as the direct compression algorithms in the sequel})
are asymptotically efficient, time critical applications ($e.g.$ pattern classification in sensing and surveillance networks) often demand
faster identification to what the state of the art can provide.
This motivates the key problem investigated in this paper: 

{\itshape Given a set of PFSA models representing patterns of interest ($i.e.$ a PFSA based pattern dictionary or library),
the problem is to identify (in real or near-real time) if any of the specified patterns of interest exist in an observed symbol sequence,
without resorting to direct compression and subsequent comparison~\cite{CR08} of  the constructed PFSA model  against the library elements}.

We propose a novel classification technique based on selective erasure of observed symbols leading to perfect information annihilation as illustrated in  Figure~\ref{figconcept1}. Specifically, we construct an additive abelian group over a slightly restricted subset of all PFSA (over a fixed alphabet),  and show that it is possible to
define  pattern-specific \textit{semantic annihilators} as a function of the group inverses. These annihilators can  then  operate on the sensed stream in a symbol-by-symbol fashion attempting to eliminate all inter-symbol correlations. The annihilation is shown to be 
perfect if and only if  the annihilator corresponds exactly to the \textit{inverted} PFSA model of the underlying generating process. Thus we need to only check if the annihilated stream (corresponding to a particular PFSA) is free from any emergent pattern, $i.e.$, if the symbols are equi-probable in an history-independent manner (denoted as \textit{symbolic white noise} in the sequel) to infer the existence of that pattern in the original observed stream.

The proposed approach is computationally efficient for direct compression of the symbol stream, since it is significantly easier to check if a symbolic stream is in fact symbolic white because
the underlying PFSA model has a single state with equal symbol generation probabilities as seen in Figure~\ref{figsimTb} and \ref{figsimTc}.  It is also shown that the proposed technique is provably faster if the cardinality of the alphabet is not greater than the number of states in a particular pattern of interest, \textit{which represents almost all PFSA models encountered in practice.}

The rest of the paper is organized in additional ten sections. Section~\ref{sec:preliminaries} is a brief overview of preliminary concepts, and related work. Section~\ref{sec:measure} presents
the construction of the additive abelian group for probability measures on symbol strings which is then shown to induce an abelian group on a restricted set of PFSA.
Section~\ref{secmc} develops a practical implementation of the PFSA sum which is then used to formulate the notion of the semantic annihilators in Section~\ref{secAnn}.
Section~\ref{secperf} identifies the theoretical conditions under which we can  guarantee  classification via semantic annihilation to be  faster than  direct compression. Section~\ref{seccomplex} establishes asymptotic bounds on the run-time complexity of annihilators.
Simulation results are presented in Section~\ref{secvv}, and pertinent discussions, intuitive interpretations, and potential applications are delineated in Section~\ref{secdiscuss}.  The paper is concluded in Section~\ref{secsumm} with recommendations for future work.

\vspace{0pt}
\section{Preliminary Concepts and Related Work} \label{sec:preliminaries}
A string $x$ over an alphabet ($i.e.$ a non-empty finite set) $\Sigma$ is a finite-length string of symbols in $\Sigma$~\cite{HMU01}. The length of a string $x$ is the number of symbols in $x$ and is denoted by $|x|$. The Kleene closure of $\Sigma$, denoted by $\Sigma^\star$, is the set of all finite-length strings of symbols including the null string $\epsilon$. The set of all strictly infinite-length strings of symbols is denoted as $\Sigma^{\omega}$. The string $xy$ is the  concatenation of strings $x$ and $y$. Therefore, the null string $\epsilon$ is the identity element of the concatenative monoid.
\begin{defn}[PFSA]
A probabilistic finite state automaton (PFSA) is a tuple $G=(Q,\Sigma,\delta,q_0,\widetilde{\Pi})$, where
 $Q$ is a (nonempty) finite set, called the set of states;
$\Sigma$ is a (nonempty) finite set, called the input alphabet;
$\delta: Q\times \Sigma\rightarrow Q$ is the state transition function;
$q_0\in Q$ is the start state;
$\widetilde{\Pi}: Q\times \Sigma \rightarrow [0,1]$ is an output mapping, known as the probability morph function that
specifies the state-specific symbol generation probabilities, and satisfies
$\forall q_i \in Q, \sigma \in \Sigma, \widetilde{\Pi}(q_i,\sigma) \geqq 0$, and $ \sum_{\tau\in\Sigma}\widetilde{\Pi}(q_i,\tau)=1$.
\end{defn}

\begin{notn}\label{notnpitpi}
 In the sequel, we would often use a matrix representation $\widetilde{\Pi}$ (denoted as the morph matrix) of the morph function, with the $ij^{th}$ element given by
$ \widetilde{\Pi}(q_i,\sigma_j)$. Note,
that $\widetilde{\Pi}$ is, in general, a rectangular non-negative matrix with row sums equal to unity. Also, from a knowledge of the morph matrix
$\widetilde{\Pi}$, and the transition map $\delta$, one can compute the stochastic state transition matrix $\Pi$, as:
\begin{gather}
 \Pi_{ij} = \sum_{\sigma_k : \delta(q_i,\sigma_k) = q_j} \widetilde{\Pi}(q_i,\sigma_k)
\end{gather}
Note that $\Pi$ is a square non-negative stochastic matrix.
\end{notn}

\begin{notn}
The transition map $\delta$ naturally induces an extended transition function $\delta^\star:Q\times\Sigma^\star\rightarrow Q$ such that $\delta^\star(q,\epsilon)=q$ and $\delta^\star(q,x\tau)=\delta(\delta^\star(q,x),\tau)$ for $q\in Q$, $x\in\Sigma^\star$ and $\tau\in\Sigma$.
\end{notn}

We assume that the underlying graph for the  PFSA models considered in this paper is irreducible, $i.e.$, is strongly connected.
This implies that the transition probability matrix $\Pi$ is an irreducible stochastic matrix, and in particular, has an unique stationary distribution~\cite{BP94} irrespective of the
the initial distribution. This assumption is motivated by the association of PFSA with emerging patterns in statistically stationary symbolic streams, because it makes little sense to represent
such dynamical systems with models whose stationary behavior would depend on the initial state. Furthermore, the theoretical development in the sequel, necessitates this assumption for technical reasons.

\begin{notn}\label{notCw}
In the sequel, we denote the PFSA constructed by directly compressing a symbol string $\omega\in \Sigma^\star$  as $\mathbb{C}(\omega)$.
The specific algorithm used is not important for the analysis presented in this paper.
\end{notn}
%
%
\begin{defn}[$\sigma$-Algebra]
A collection $\mathfrak{M}$ of subsets of a non-empty set $X$ is
said to be a $\sigma$-algebra~\cite{R88} in $X$ if $\mathfrak{M}$
has the following properties:
\begin{enumerate}
\item $X \in \mathfrak{M}$ \item If $A \in \mathfrak{M}$, then
$A^c \in \mathfrak{M}$ where $A^c$ is the complement of $A$
relative to $X$, i.e., $A^c = X\setminus A$ \item If $A =
\bigcup_{n= 1}^{\infty} A_n$ and if $A_n \in \mathfrak{M}$ for
$n\in \mathbb{N}$, then $A \in \mathfrak{M}$.
\end{enumerate}
\end{defn}
\begin{defn} [Measure] A (non-negative) measure is a  countably
additive function $\mu$, defined on a $\sigma$-algebra
$\mathfrak{M}$, whose range is $[0, \infty]$. Countable additivity means that if $\{A_i\}$ is a pairwise
disjoint countable collection of members of $\mathfrak{M}$, then
$\mu \left (\bigcup_{i=1}^\infty A_i \right ) = \sum_{i=1}^\infty \mu (A_i)$
\end{defn}
\begin{defn}[Probability Measure]
A probability measure on a non-empty set with a specified
$\sigma$-algebra $\mathfrak{M}$ is a finite (non-negative) measure
on $\mathfrak{M}$. Although not required by the theory, a
probability measure is  defined to have the unit interval $[0,1]$
as its range.
\end{defn}
%
\begin{figure}[t]
\centering
\subfloat[]{\begin{minipage}{2.3in}
\psfrag{N}[lc]{ \Mblue Nerode equivalent strings}
\psfrag{S}[cc]{ \bf States}
\psfrag{w1}[lc]{$\omega_1$}
\psfrag{w2}[lc]{$\omega_2$}
\psfrag{w3}[cc]{$\omega_3$}
\psfrag{F}[lc]{  \Dgreen \scshape Future}
\psfrag{P}[cl]{  \Darkred \scshape Past}
\psfrag{q1}[cc]{\Large $\phantom{.}q_1$}
\psfrag{q2}[cc]{\Large $\phantom{.}q_2$}
 \includegraphics[width=3in]{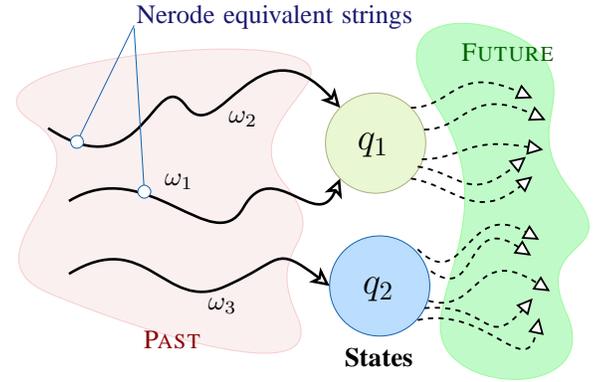}\end{minipage}\label{figstate}
}
\hfill
\subfloat[]{\includegraphics[width=2.5in]{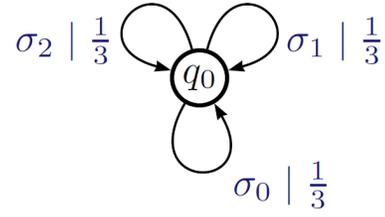}\label{figsimTb}}\\
\subfloat[]{\includegraphics[width=2in]{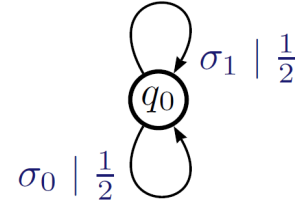}\label{figsimTc}}
\caption{Linguistic Concepts: (a) Concept of PFSA states from Probabilistic Nerode Equivalence: Nerode equivalent strings $\omega_1,\omega_2$
 have probabilistically indistinguishable future evolution, thus leading to the same state $q_1$. (b) Symbolic White Noise (See Eq.~\eqref{eqwn} for formal definition) for alphabet $\Sigma = \{\sigma_0,\sigma_1,\sigma_2\}$; (c) for alphabet $\Sigma = \{\sigma_0,\sigma_1\}$}
%
\end{figure}
\begin{defn}[Measure Space]
A probability measure space is a triple $(X,\mathfrak{M},p)$
where $X$ is a non-empty set, $\mathfrak{M}$ is a
$\sigma$-algebra in $X$, and $p$ is a finite non-negative measure
on $\mathfrak{M}$.
\end{defn}
\begin{defn}[$\sigma$-Algebra for Symbolic Strings]
Given an alphabet $\Sigma$, the set $\mathfrak{B}_\Sigma \triangleq
2^{\Sigma^\star}\Sigma^\omega$ is defined to be the
$\sigma$-algebra generated by the set $ \{ L : L =
x\Sigma^\omega \ \mathrm{where} \ x \in \Sigma^\star
 \}$, i.e., the smallest $\sigma$-algebra on the set
$\Sigma^\omega$, which  contains the set $ \{ L : L =
x\Sigma^\omega \ \mathrm{s.t.} \ x
\in \Sigma^\star \}$.
\end{defn}

For brevity, the probability $p(x\Sigma^{\omega})$ is denoted as $p(x),\forall x\in\Sigma^\star$ in the sequel. In other words, $p(x)$ is the probability of the occurrence of all the strings with $x$ as a prefix.
\begin{defn}[Probabilistic Nerode Relation]
Given an alphabet $\Sigma$, any two strings $x,y\in\Sigma^\star$ are said to satisfy the probabilistic Nerode relation $\mathcal{N}_p$ on a probability space $(\Sigma^{\omega},\mathfrak{B}_\Sigma,p)$, denoted by $x\mathcal{N}_py$, if either of the following conditions is true:
\begin{enumerate}
 \item $ p(x)=p(y)=0$;
 \item $\forall \sigma\in\Sigma^\star,\frac{p(x\sigma)}{p(x)}=\frac{p(y\sigma)}{p(y)}$ provided that $p(x)\neq 0 ,p(y)\neq 0$.

\end{enumerate}
\end{defn}
\vspace{4pt}
It has been proved in~\cite{CR08} that the probabilistic Nerode relation defined above is a right-invariant equivalence relation~\cite{HMU01}
which means that if two strings $x,y$ are equivalent, so are any right extensions of the strings, $i.e.$,
\begin{gather*}
 \forall x,y,u \in \Sigma^\star, \ x\mathcal{N}_py \Rightarrow xu\mathcal{N}_pyu
\end{gather*}
 In the sequel, this is referred to as probabilistic Nerode equivalence and we denote the Nerode equivalence class of a string $x$ on $\Sigma^\star$ by $[x]_p$, i.e., $[x]_p=\{z\in\Sigma^\star: x\mathcal{N}_p z\}$. The right invariance property induces the notion of states and hence is crucial to the definition of probabilistic state machines;
 by this property two equivalent strings have probabilistically indistinguishable future
evolution and therefore can be visualized as terminating on the same state as seen in Figure~\ref{figstate}. In this context, we make the following observation:
\begin{quote}
A symbolic dynamical process has a probabilistic finite state description if and only if the corresponding Nerode equivalence has a finite index.
\end{quote}
\begin{defn}[Space of PFSA]\label{def:allmacandmeas}
The space of all PFSA over a given symbol alphabet is denoted by $\mathscr{A}$ and the space of all probability measures  $p$
that induce a finite-index probabilistic Nerode equivalence on the corresponding measure space $(\Sigma^{\omega},\mathfrak{B}_\Sigma,p)$
is denoted by $\Q$.
\end{defn}
As expected, there is a close relationship between $\mathscr{A}$ and $\Q$, which is made explicit in the sequel.
\begin{defn}[PFSA Map $\mathds{H}$] \label{def:H}
Let $p\in\mathscr{P}$ and $G=(Q,\Sigma,\delta,q_0,\widetilde{\Pi})\in\mathscr{A}$. The map $\mathds{H}:\mathscr{A}\rightarrow\mathscr{P}$ is defined as $\mathds{H}(G)=p$ such that the following condition is satisfied:
\begin{gather*} \label{equ:H}
\forall x = \sigma_1\cdots\sigma_r\in\Sigma^\star, \\ 
p(x)=\widetilde{\Pi}(q_0,\sigma_1) \prod_{k=1}^{r-1} \widetilde{\Pi}(\delta^\star(q_0,\sigma_1\cdots\sigma_k),\sigma_{k+1})
\end{gather*}
 where
$r\in\mathbb{N}$, the set of positive integers.
\end{defn}
\begin{defn}[Right Inverse $\mathds{H}_{-1}$]
The right inverse of the map $\mathds{H}$ is denoted by $\mathds{H}_{-1}:\mathscr{P}\rightarrow\mathscr{A}$ such that
\begin{gather*}
 \forall p \in \Q, \ \mathds{H}(\mathds{H}_{-1} (p)) = p
\end{gather*}
\end{defn}

An explicit construction of the map $\mathds{H}_{-1}$ is reported in~\cite{CR08} and is not presented in this paper, because we only require that such a map exists.
\begin{defn}[Perfect Encoding]\label{def:encodingM}
Given an alphabet $\Sigma$, a PFSA $G=(Q,\Sigma,\delta,q_0,\widetilde{\Pi})$ is said to be a perfect encoding of the measure space $(\Sigma^{\omega},\mathscr{B}_{\Sigma},p)$ if $p=\mathds{H}(G)$.
\end{defn}

There are possibly many PFSA realizations that encode the same probability measure on $\mathscr{B}_{\Sigma}$ due to existence of non-minimal realizations and state relabeling; neither of them affect the underlying encoded measure. From this perspective, a notion of PFSA equivalence is introduced as follows:
\begin{defn}[PFSA Equivalence]\label{def:equPFSA}
 Two PFSA $G_1$ and $G_2$ are defined to be equivalent if $\mathds{H}(G_1) = \mathds{H}(G_2)$. In this case, we say $G_1 = G_2$.
\end{defn}
\begin{rem}
In the sequel, a PFSA $G$ implies the equivalence class of $G$, $i.e.$,   $\{P\in\mathscr{A}:\mathds{H}(P) = \mathds{H}(G)\}$. This concept is similar to the equivalence class of almost everywhere equal functions being a unique vector in the $L_r$-space~\cite{R88}.
\end{rem}
\begin{defn}[Structural Equivalence] \label{def:sameStruct}
Two PFSA $G_i=(Q_i,\Sigma,\delta_i,q_0^i,\widetilde{\Pi}_i)\in\mathscr{A}$, $i=1,2$, are defined to have the equivalent (or identical) structure if $Q_1=Q_2,q_0^1 = q_0^2$ and $\delta_1(q,\sigma)=\delta_2(q,\sigma),\forall q\in Q_1 \ \forall\sigma\in\Sigma$.
\end{defn}
\begin{defn}[Synchronous Composition of PFSA] \label{def:product}
The binary operation of synchronous composition of two PFSA $G_i=(Q_i,\Sigma,\delta,q_0^{(i)},\widetilde{\Pi}_i)\in\mathscr{A}$ where $i=1,2$, denoted by $\sync: \mathscr{A}\times\mathscr{A}\rightarrow\mathscr{A}$ is defined as 
\begin{gather*}
G_1 \sync G_2 = \left  (Q_1\times Q_2,\Sigma,\delta',(q_0^{(1)},q_0^{(2)}),\widetilde{\Pi}'\right ) 
\end{gather*}
 where $\delta'$ and $\widetilde{\Pi}'$ is computed as follows:
\begin{gather*}
 \forall q_i\in Q_1,q_j\in Q_2,\sigma\in\Sigma,\\  \left \{ \begin{array}{l} \delta' ( (q_i,q_j),\sigma) = \left( \delta_1(q_i,\sigma),\delta_2(q_j,\sigma) \right)\\
 \widetilde{\Pi}'((q_i,q_j),\sigma)=\widetilde{\Pi}_1 (q_i,\sigma)
\end{array}\right.
\end{gather*}
\end{defn}
\vspace{4pt}
\begin{rem}
In general, the operation $\sync$ of synchronous composition is non-commutative.
\end{rem}
\begin{prop}[Synchronous Composition of PFSA] \label{pro:productSameMeasure}
Let $G_1, G_2\in\mathscr{A}$. Then,
 $\mathds{H}(G_1)=\mathds{H}(G_1 \sync G_2)$ and therefore $G_1=G_1 \sync G_2$ in the sense of Definition~\ref{def:equPFSA}.
\end{prop}
\begin{IEEEproof}
See Theorem 4.5 in~\cite{CR08}.
\end{IEEEproof}

Synchronous composition of PFSA allows transformation of PFSA with disparate structures to non-minimal descriptions that have the
same underlying graphs. This assertion is crucial for the development in the sequel, since any binary operation defined for two
PFSA with an identical structure can be extended to the general case on account of Definition~\ref{def:product} and Proposition~\ref{pro:productSameMeasure}.
\vspace{0pt}
\section{Abelian Group of PFSA} \label{sec:measure}
This section shows that a subspace of PFSA can be assigned the algebraic structure of an abelian group.
We first construct the abelian group on a subspace of probability measures, and then induce the group structure
on this subspace of PFSA via the isomorphism between the two spaces.
\begin{defn}[Restricted PFSA Space]\label{defnA+}
Let $\mathscr{A}^+=\{ G=(Q,\Sigma,\delta,q_0,\widetilde{\Pi}): \widetilde{\Pi}(q,\sigma)>0 \ \forall q\in Q \ \forall \sigma\in\Sigma\}$ that is a proper subset of $\mathscr{A}$. It follows that the transition map of any PFSA in the subset $\mathscr{A}^+$ is a total function. We restrict the map $\mathds{H}:\mathscr{A}\rightarrow\mathscr{P}$ on a smaller domain $\mathscr{A}^+$, that is, $\mathds{H}^+:\mathscr{A}^+\rightarrow\mathscr{P}^+$, i.e., $\mathds{H}^+=\mathds{H}\arrowvert_{\mathscr{A}^+}$.
\end{defn}
\begin{defn}[Restricted Probability Measure]\label{defnP+}
Let $\mathscr{P}^+\triangleq\{p\in\mathscr{P}:p(x)\neq 0,\forall x\in\Sigma^\star\}$ that is a proper subset of $\mathscr{P}$. Each element of $\mathscr{P}^+$ is a probability measure that assigns a non-zero probability to each string on $\mathfrak{B}_\Sigma$.
Similar to Definition~\ref{defnA+}, we  restrict $\mathds{H}_{-1}$ on  $\mathscr{P}^+$, i.e., $\mathds{H}_{-1}^+=\mathds{H}_{-1}\arrowvert_{\mathscr{P}^+}$.
\end{defn}

Since we do not distinguish PFSA in the same equivalence class (See Definition~\ref{def:equPFSA}), we have the following result.
\begin{prop}[Isomorphism of $\mathds{H}^+$] \label{pro:isom}
The map $\mathds{H}^+$ is an isomorphism between the spaces $\mathscr{A}^+$ and $\Q^+$, and its inverse is $\mathds{H}^+_{-1}$.
\end{prop}
%
\begin{defn}[Abelian Operation on $\mathscr{P^+}$] \label{def:addition}
 The addition operation $\oplus: \mathscr{P^+} \times \mathscr{P^+} \rightarrow \mathscr{P^+}$ is defined by $p_3\triangleq p_1 \oplus p_2, \forall p_1,p_2 \in \mathscr{P^+}$ such that
\begin{enumerate}
  \item $p_3(\epsilon)=1$.
  \item $\forall x \in \Sigma^\star$ and $\tau\in\Sigma$, $\frac{p_3(x\tau)}{p_3(x)}=\frac{p_1(x\tau)p_2(x\tau)}{\sum_{\alpha\in\Sigma} p_1(x\alpha)p_2(x\alpha)}$
 \end{enumerate}
\end{defn}
\vspace{3pt}
$p_3$ is a well-defined probability measure on $\mathscr{P^+}$, since
\begin{gather*}
 \forall x\in\Sigma^\star,\\ \Sigma_{\tau\in\Sigma}p_3(x\tau)= \Sigma_{\tau\in\Sigma}\frac{p_1(x\tau)p_2(x\tau)}{\sum_{\alpha\in\Sigma} p_1(x\alpha)p_2(x\alpha)}p_3(x)= p_3(x)
\end{gather*}
\begin{prop}[abelian Group of PFSA]\label{prop:abelian}
 The algebra $(\mathscr{P^+},\oplus)$ forms an abelian group.
\end{prop}

\vspace{4pt}\begin{IEEEproof}
Closure property and commutativity of $(\mathscr{P^+},\oplus)$ are obvious. The associativity, existence of identity and existence of inverse element are established next.\\
\textit{(1) Associativity} $i.e.$ $(p_1\oplus p_2)\oplus p_3 = p_1\oplus (p_2\oplus p_3)$. We note, that $\forall x\in\Sigma^\star,\tau\in\Sigma$,
\begin{align*}
    \frac{((p_1\oplus p_2)\oplus p_3)(x\tau)}{((p_1\oplus p_2)\oplus p_3)(x)}
     =& 
\frac{(p_1\oplus p_2)(x\tau)p_3(x\tau)}{\sum_{\beta\in\Sigma} (p_1\oplus p_2)(x\beta)p_3(x\beta)}
\\
    = &
\frac{\frac{p_1(x\tau)p_2(x\tau)}{\sum_{\alpha\in\Sigma}p_1(x\alpha)p_2(x\alpha)}p_3(x\tau)}{\sum_{\beta\in\Sigma} \frac{p_1(x\beta)p_2(x\beta)}{\sum_{\alpha\in\Sigma} p_1(x\alpha)p_2(x\alpha)}p_3(x\beta)} 
\notag\\
     =
     &
     \frac{p_1(x\tau)p_2(x\tau)p_3(x\tau)}{\sum_{\beta\in\Sigma}p_1(x\beta)p_2(x\beta)p_3(x\beta)}
\\
     =&
\frac{p_1(x\tau) \frac{p_2(x\tau)p_3(x\tau)}{\sum_{\alpha\in\Sigma} p_2(x\alpha)p_3(x\alpha)}}{\sum_{\beta\in\Sigma} p_1(x\beta)\frac{p_2(x\beta)p_3(x\beta)}{\sum_{\alpha\in\Sigma} p_2(x\alpha)p_3(x\alpha)}}  
\\
 =&
\frac{p_1(x\tau) (p_2\oplus p_3)(x\tau)}{\sum_{\beta\in\Sigma} p_1(x\beta)(p_2\oplus p_3)(x\beta)}     
\notag \\
&
=
\frac{(p_1\oplus (p_2\oplus p_3))(x\tau)}{(p_1\oplus (p_2\oplus p_3))(x)}
\end{align*}
\textit{ (2) Existence of identity}: Let us introduce a probability measure $\mathbf{i}_\circ$ of symbol
strings such that:
\begin{gather}\label{eqwn}
\forall x\in\Sigma^\star, \ \mathbf{i}_\circ(x)=\left(\frac{1}{|\Sigma|}\right)^{|x|}
\end{gather}
where $|x|$ denotes the length of the string $x$. Then, $\forall\tau\in\Sigma$ that $\frac{\mathbf{i}_\circ(x\tau)}{\mathbf{i}_\circ(x)}=\frac{1}{|\Sigma|}$. For a measure $p\in\mathscr{P^+}$ and $\forall \tau\in \Sigma$,
\begin{align*}
\frac{(p\oplus \mathbf{i}_\circ)(x\tau)}{(p\oplus \mathbf{i}_\circ)(x)} & = \frac{p(x\tau)\mathbf{i}_\circ(x\tau)}{\sum_{\alpha\in\Sigma}p(x\alpha)\mathbf{i}_\circ(x\alpha)}
=\frac{p(x\tau)}{p(x)}
\end{align*}
This implies that $p\oplus \mathbf{i}_\circ = \mathbf{i}_\circ\oplus p = p$ by
Definition \ref{def:addition} and by commutativity.  Therefore,
$\mathbf{i}_\circ$ is the identity of the monoid $(\mathscr{P}^+, \oplus)$. \\
\textit{(3) Existence of inverse}: $\forall p\in\mathscr{P^+}$, $\forall x\in\Sigma^\star$ and $\forall \tau\in\Sigma$, let $-p$ be defined by the following relations:
\begin{align}
     (-p)(\epsilon)&=1 \\
    \frac{(-p)(x\tau)}{(-p)(x)}&=\frac{p^{-1}(x\tau)}{\sum_{\alpha\in\Sigma}p^{-1}(x\alpha)}
    \intertext{Then, we have:}
\frac{(p\oplus (-p))(x\tau)}{(p\oplus (-p))(x)}
     & =\frac{p(x\tau)(-p)(x\tau)}{\sum_{\alpha\in \Sigma}p(x\alpha)(-p)(x\alpha)}
    =\frac{1}{|\Sigma|}
    \end{align}
This gives $p\oplus (-p)=\mathbf{i}_\circ$ which completes the proof.
\end{IEEEproof}
In the sequel, we denote  the zero-element $\mathbf{i}_\circ$ of the abelian group $(\mathscr{P}^+, \oplus)$ as the \emph{symbolic white noise}. The concept of symbolic white noise has been illustrated in Figure~\ref{figsimTb} and \ref{figsimTc}.
\vspace{0pt}
\subsection{Explicit Computation of the abelian Operation $\oplus$} \label{sec:relationship}

The isomorphism between $\mathscr{P}^+$ and $\mathscr{A}^+$ (See Proposition~\ref{pro:isom})
induces the following abelian operation on
$\mathscr{A}^+$.
\begin{defn}[Addition Operation on PFSA] \label{def:PFSAoperation}
Given any $G_1,G_2\in\mathscr{P}^+$, the addition operation $\pfsum :\mathscr{A}^+\times\mathscr{A}^+\rightarrow\mathscr{A}^+$ is defined as: $$G_1 \ \pfsum \ G_2=\mathds{H}_{-1}^+(\mathds{H}^+(G_1)\oplus\mathds{H}^+(G_2))$$
\end{defn}

If the summand PFSA have identical structure (i.e., their underlying graphs are identical), then the
explicit computation of this sum is stated as follows.
\begin{prop} [PFSA Addition]\label{pro:sameStructure}
 If two PFSA $G_1,G_2\in\mathscr{A}^+$ are of the same structure, i.e., $G_i=(Q,\Sigma,\delta,q_0,\widetilde{\Pi}_i),i=\{1,2\}$, then we have $G_1 \pfsum G_2 = (Q,\Sigma,\delta,q_0, \widetilde{\Pi})$ where
 \begin{gather}
   \widetilde{\Pi}(q,\sigma)=\frac{\widetilde{\Pi}_1(q,\sigma)\widetilde{\Pi}_2(q,\sigma)}{\sum_{\alpha\in\Sigma}\widetilde{\Pi}_1(q,\alpha)\widetilde{\Pi}_2(q,\alpha)}
  \end{gather}
\end{prop}
\vspace{5pt}
\vspace{4pt}\begin{IEEEproof}
 Let $p_i=\mathds{H}^+(G_i)$, $i=\{1,2\}$ and since $G_1,G_2$ have the same structure, we have from  Eq. \eqref{equ:H}:
\begin{gather}
\forall \sigma \in \Sigma, \forall x \ \mathrm{s.t.} \  \delta^\star(q_0,x) = q \in Q, \notag \\ \frac{p_i(x\sigma)}{p_i(x)}=\widetilde{\Pi}_i(\delta^\star(q_0,x),\sigma)=\widetilde{\Pi}_i(q,\sigma)
\end{gather}
Now, by Definition \ref{def:addition} and Definition \ref{def:H},
\begin{align*}
& \quad \widetilde{\Pi}(q,\sigma)
=\frac{(p_1\oplus p_2)(x\sigma)}{(p_1\oplus p_2)(x)}
=\frac{p_1(x\sigma)p_2(x\sigma)}{\sum_{\alpha\in\Sigma}p_1(x\alpha)p_2(x\alpha)}\\ =&\frac{\frac{p_1(x\sigma)p_2(x\sigma)}{p_1(x)p_2(x)}}{\sum_{\alpha\in\Sigma}\frac{p_1(x\alpha)p_2(x\alpha)}{p_1(x)p_2(x)}}
=\frac{\widetilde{\Pi}_1(q,\sigma)\widetilde{\Pi}_2(q,\sigma)}{\sum_{\alpha\in\Sigma}\widetilde{\Pi}_1(q,\alpha)\widetilde{\Pi}_2(q,\alpha)}
\end{align*}
\end{IEEEproof}
%
The extension to the general case is achieved by using synchronous composition of probabilistic machines.
\begin{prop}[PFSA Addition (General case)]\label{propPsum}
Given two PFSA $G_1, G_2\in\mathscr{A}^+$, the sum
$G_1 \pfsum  G_2$ is computed via Proposition  \ref{pro:sameStructure} and Definition  \ref{def:product} as follows:
       \begin{equation}
        G_1  \pfsum  G_2 = (G_1 \sync G_2) \pfsum ( G_2 \sync G_1)
       \end{equation}
\end{prop}
\begin{IEEEproof}
Noting that $G_1\sync G_2$ and $G_2\sync G_1$ have the same structure up to state relabeling, it follows from Proposition~\ref{pro:productSameMeasure}:
\begin{align*}
 \mathds{H}^+(G_1  \pfsum  G_2)
 &=\mathds{H}^+(G_1)\oplus\mathds{H}^+(G_2) \mspace{30mu} \mathrm{(See\ Definition \ \ref{def:PFSAoperation})}\\
& =\mathds{H}^+(G_1 \sync G_2)\oplus\mathds{H}^+(G_2 \sync G_1)\\ &=\mathds{H}^+\bigg ((G_1 \sync G_2)  \pfsum  (G_2 \sync G_1)\bigg )
 \end{align*}
which completes the proof.
\end{IEEEproof}
\begin{example}
Let $G_1$ and $G_2$ be two PFSA with identical structures, such that the probability morph matrices are:
\begin{gather}
 \widetilde{\Pi}_1 =
\begin{pmatrix}
 0.2 & 0.8 \\
 0.4 & 0.6
\end{pmatrix}
\textrm{  and }
\widetilde{\Pi}_2 =
\begin{pmatrix}
 0.1 & 0.9 \\
 0.6 & 0.4
\end{pmatrix}
\end{gather}
Then the $\widetilde{\Pi}$-matrix for the sum $G_1 \pfsum G_2$, denoted by $\widetilde{\Pi}_{12}$, is
\begin{gather*}
\widetilde{\Pi}_{12} =
\begin{pmatrix}
 \Mblue0.1\times0.2 & \Mblue0.9\times 0.8 \\
 \Mblue0.6 \times 0.4 & \Mblue0.4 \times 0.6
\end{pmatrix} \xrightarrow[{\Dgreen rows}]{{\Dgreen Normalize}}
\begin{pmatrix}
 0.027 & 0.973 \\
 0.5 & 0.5
\end{pmatrix}
\end{gather*}
\end{example}
\vspace{0pt}
\section{ A Machine Representation of PFSA Sum}\label{secmc}
In this section, we investigate the
implementation of the sum of two PFSA by a sequentially controlled interaction of individually
generated symbol strings, which form the conceptual basis of designing a semantic annihilator.
Referring to Figure~\ref{fig1}, we will call this the {\sffamily Plus}-machine.
%
\begin{figure}[!ht]
\centering
\psfrag{G}[cc]{$G_1$}
\psfrag{H}[cc]{$G_2$}
\psfrag{A}[cc]{\scriptsize \sffamily\scshape{$\phantom{.}$and}}
\psfrag{J}[cr]{\scriptsize \Dblue Switch}
\psfrag{O}[tl]{\Mblue \txt{Output\\ Sequence}}
\psfrag{S}[cl]{\BRed$G_1 \pfsum G_2$}
\psfrag{E1}[tl]{}
\psfrag{E2}[bc]{}
 \includegraphics[width=2.25in]{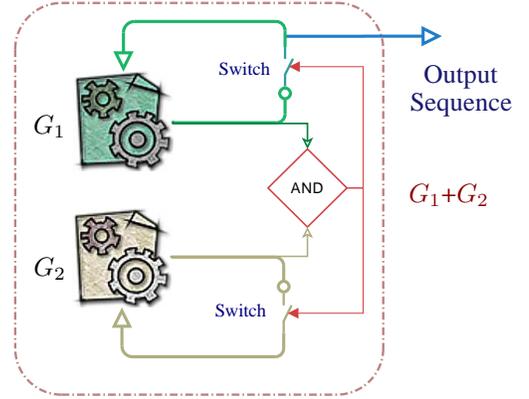}
\caption{Sum of two PFSA: The {\sffamily Plus}-machine $\mathscr{M}(G_1 \pfsum G_2)$. The machines generate symbols independently, but is allowed to change states only if the generated symbols match.}\label{fig1}
\vspace{0pt}
\end{figure}
\vspace{0pt}
\subsection{Functional Description of the {\sffamily Plus}-Machine}
For a given pair of PFSA $G_1$ and $G_2$, the {\sffamily Plus}-machine denoted as as $\mathscr{M}$$
(G_1 \pfsum G_2)$ has the following components:
\begin{itemize}
 \item Copies of the component machines $G_1$ and $G_2$. We assume, without loss of generality, that $G_1$ and $G_2$ have the same structure (See Definition~\ref{def:product}), since
this can be always arranged (See Proposition~\ref{pro:productSameMeasure}).
\item A logical $\and$ gate $\and:\Sigma \times \Sigma \rightarrow \{0,1\}$ which operates as follows:
\begin{gather*}
 \forall \sigma_i,\sigma_j \in \Sigma, \ \sigma_i \ \and \ \sigma_j = \left \{ \begin{array}{ll}
                                       0 \ \textrm{(false)},& \mathrm{if} \ \sigma_i \neq \sigma_j \\
                    1 \ \textrm{(true)},& \mathrm{otherwise}
                                      \end{array}\right.
\end{gather*}
\end{itemize}
\vspace{0pt}
\subsection{Operational Description of the {\sffamily Plus}-Machine}
The $\pfsum$ machine $\mathscr{M}(G_1 \pfsum G_2)$ operates as follows:
\begin{itemize}
\item Each of the component machines, $G_1$ and $G_2$, is initialized to the same state $q_0$ in the
underlying graph.
\item Each of the component machines, $G_1$ and $G_2$, operates in a statistically  independent manner  to generate symbols from the alphabet $\Sigma$.
\item However, to activate a state transition, the generated symbols must be passed through the $\and$ gate, upon which they must yield a true output. Formally,
\begin{align*}
&\forall \sigma_i,\sigma_j \in \Sigma, q_i, q_j \in Q, \\ & \delta\bigg ((q_i,q_j)(\sigma_i,\sigma_j)\bigg ) = \left \{ \begin{array}{lll}
                                               (q_i,q_j), & \\ &\mspace{-100mu}\mathrm{if} \ \sigma_i \ \and \ \sigma_j = 0 & \\
                        \bigg (\delta(q_i,\sigma_i),\delta(q_j,\sigma_j)\bigg), & \\ & \mspace{-100mu}\mathrm{otherwise} &
                                              \end{array}\right.
\end{align*}
\item The machine is assumed to function inside a ``black box'', with an external observer. The observable output string generated as follows: \textit{A generated symbol is
observable if and only  if it causes a sMargtate transition}.
\end{itemize}
The sequential functioning of $\mathscr{M}(G_1 \pfsum G_2)$ is illustrated in Figure~\ref{fig1}. We have the following result:
\begin{prop}[Semantic Compression]\label{prop+mc}
 For a given pair of PFSA $G_1,G_2$, if the output string from the  $\mathscr{M}(G_1 \pfsum G_2)$ is denoted as $x \in \Sigma^\omega$, then, the PFSA $\mathds{C}(x)$ obtained by semantically compressing $x$ is given by the sum $G_1 \pfsum G_2$.

\end{prop}
\vspace{4pt}\begin{IEEEproof}
 It follows from the functional description, and the following considerations:
\begin{enumerate}
 \item The component machines $G_1$ and $G_2$ are always state synchronized (follows from  operational description).
\item The components generate symbols in a statistically independent manner.
\item  The probability for $\mathscr{M}(G_1 \pfsum G_2)$ to emit a particular symbol $\sigma \in \Sigma$, while being at state $(q_i,q_i)$, ($i.e.$ both components are at state $q_i$), is given by the probability of generating $\sigma$ simultaneously (and independently) by both components; and the probability of this compound symbol (marginalized by the probability of generating identical symbols on both machines)  is :
\begin{gather*}
 \widetilde{\Pi}_{12}((q_i,q_i),(\sigma,\sigma)) \\ =\frac{\widetilde{\Pi}_1(q_i,\sigma)\widetilde{\Pi}_2(q_i,\sigma)}{\sum_\sigma \widetilde{\Pi}_1(q_i,\sigma)\widetilde{\Pi}_2(q_i,\sigma)}
\begin{array}{ll}
\Dblue\longleftarrow  \textit{\txt{Compound  Event} }\\
\Dblue \textit{\txt{$\longleftarrow$Marginalization} }
\end{array}
\end{gather*}
which matches exactly with Proposition~\ref{pro:sameStructure}.
\item Since the internal states of $\mathscr{M}(G_1 \pfsum G_2)$ are always of the form $(q_i,q_i)$, it is straightforward to see that for any correct semantic compression algorithm,
the structure of the identified PFSA matches with the component machines, $G_1$ and $G_2$. The proof is now complete.
\end{enumerate}
\end{IEEEproof}
%
It follows from Proposition~\ref{prop+mc}, that the {\sffamily Plus}-Machine can be used to annihilate information in the symbol string generated by a PFSA in the following manner:
\begin{gather}
 G \pfsum H= \Hi(\W) \Rightarrow \mathscr{M}(G \pfsum H) = \Hi(\W)
\end{gather}
which implies that if $G$ is the underlying PFSA for the sensed process, and we can compute $H$ such that $G+H= \Hi(\W)$,
and subsequently modify the incoming sensed data stream via the {\sffamily Plus}-machine construction, we would end up with symbolic white noise in the output, which
then can be identified easily. This, however, is not directly achievable in practice for the following reasons:
\begin{enumerate}
 \item Impossibility of  state synchronization with sensed stream.
\item Impossibility  of disabling state transitions in the sensed physical process.
\end{enumerate}
The next section presents modifications to this basic construction to admit a physically realizable implementation of a semantic annihilator.
\vspace{0pt}
\section{Semantic Annihilation}\label{secAnn}
In this section, we assume that we are given a pre-identified (during the training phase) pattern library $\pti \triangleq \{G^i: G^i \in \mathscr{A}^+\}$ containing a
finite number of patterns of interest, represented as PFSA. We would construct a \textit{semantic annihilator} for each pattern in $\pti$, which
would be used in online classification.

We need the following function that operates symbol-wise on streams, typically
implementing a selective erasure of the two input streams ($\epsilon$ is the null symbol, $i.e.$, the identity in the concatenative free monoid over the alphabet $\Sigma$):
\begin{defn}[Erasing Function] The erasing function $\xi:\Sigma \times \Sigma \rightarrow \Sigma \bigcup \{\epsilon\}$ is defined as follows:
\begin{gather}
 \xi(\sigma_1,\sigma_2) = \left \{ \begin{array}{cl} \sigma_1 & \mathrm{if} \ \sigma_1 \ \and \ \sigma_2 = 1 \\
                                    \epsilon & \mathrm{otherwise}
                                   \end{array}\right.
\end{gather}
\end{defn}
\subsection{Construction of the Semantic Annihilator}\label{subsecann}
\begin{figure}[!ht]
 \centering
\psfrag{S}[rb]{\txt{\bf Sensed\\ \bf Stream}}
\psfrag{E}[rc]{$\boldsymbol{\xi}$}
\psfrag{A}[cc]{\bf \footnotesize \txt{\\  \Aqua Semantic \\   \Aqua Annihilator}}
\psfrag{H1}[cc]{$\mspace{6mu}H^1$}
\psfrag{H2}[cc]{$\mspace{6mu}H^2$}
\psfrag{Hj}[cc]{$\mspace{6mu}H^m$}
\psfrag{C}[cc]{\footnotesize \txt{\bf Check for \\ \bf White Noise}}
 \includegraphics[width=3in]{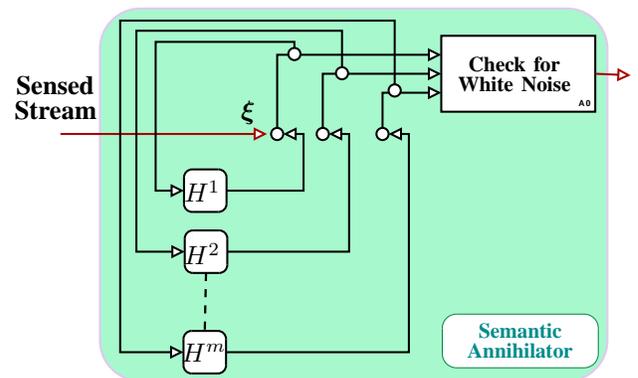}
\caption{The block design for a semantic annihilator}\label{figconcept}
\end{figure}

The component machines are set as follows:
\begin{itemize}
\item  Let the  $G\in \mathds{G}$ be one element of the pattern library.

\item Construct the additive inverse for $G$, $i.e.$ compute $H$ s.t.
\begin{gather}
 G \pfsum H = \mathds{H}^+_{-1}(\mathbf{i}_\circ)
\end{gather}
Let the state set for $H$ be $Q_H$, and let $\vert Q_H\vert = m$.
\item Create $m$ copies for $H$, each initialized at a distinct state. Let
$H^j$ be the copy of the PFSA $H$ initialized at state $q_j\in Q_H$.
\end{itemize}
\subsection{Operational Description of the Annihilator}
The semantic annihilator operates as follows:
\begin{enumerate}
 \item Read symbol $\sigma_{sensor}$ from sensor
\item Independently generate symbols $\sigma^j$ for each component $H^j$.
\item Transition each $H^j$ using the same symbol $\sigma_{sensor}$.
\item Construct $m$ symbol streams $\omega^j \in \Sigma^\star : j \in \{1,\cdots, m\}$ recursively using the erasing function $\xi$:
\begin{gather}
 \omega^j_{UPDATE} = \omega^j\xi(\sigma_{sensor},\sigma^j)
\end{gather}
\item Check if any  $\omega^j$ is in fact symbolic  white noise.
\end{enumerate}

Next we present the  main result (Proposition~\ref{propmain}) which rigorously establishes the annihilation concept as a viable tool for pattern classification.

\begin{prop}[Main Result]\label{propmain}
 At least one of the constructed streams $\omega^j$ will semantically compress to symbolic white noise
 if and only if $G \pfsum H = \mathds{H}^+_{-1}(\mathbf{i}_\circ)$, $i.e.$,
\begin{gather}
 G \pfsum H^j = \mathds{H}^+_{-1}(\mathbf{i}_\circ) \Leftrightarrow \exists j \ \big ( \mathbb{C}(\omega^j) = \Hi(\W) \big )
\end{gather}
(See Notation \ref{notCw}, and note that $H^j$ is a copy of $H$ initialized at the $j^{th} $ state. )
\end{prop}
\vspace{2pt}
\vspace{4pt}\begin{IEEEproof}\textbf{(Left to Right:)}
Let the sensed process be  generated by the underlying PFSA $G$, such that $G \pfsum H =\Hi(\W)$. We note that, by construction, there exists $j_\star$ such that
$H^{j_\star}$ is always state synchronized with $G$. However, we only see symbols in the output  stream $\omega^{j_\star}$, if the generated symbols are identical. It follows
that, on compression $\omega^{j_\star}$ would yield a modified PFSA (denote by $G^{mod}$) with structure identical to $G$, but each row of the $\widetilde{\Pi}^{mod}$ matrix would be
modified as follows:
\begin{gather*}
 \widetilde{\Pi}^{mod}(q_i,\sigma) = \frac{\widetilde{\Pi}^G(q_i,\sigma)\widetilde{\Pi}^H(q_i,\sigma)}{\sum_\sigma\widetilde{\Pi}^G(q_i,\sigma)\widetilde{\Pi}^H(q_i,\sigma)}\\
= \frac{\widetilde{\Pi}^G(q_i,\sigma)(\widetilde{\Pi}^G(q_i,\sigma))^{-1} K(q_i)}{\sum_\sigma\widetilde{\Pi}^G(q_i,\sigma)(\widetilde{\Pi}^G(q_i,\sigma))^{-1} K(q_i)}
= \frac{1}{\vert \Sigma \vert}
\end{gather*}
where $K(q_i) = \sum_\sigma (\widetilde{\Pi}^G(q_i,\sigma))^{-1}$ is the normalizing constant, implying each row is identical and uniform which in turn implies that the identified model is symbolic white noise.\\
\textbf{(Right to Left:)} We show this by contradiction as follows: Let  the sensed process is generated by $G$ such that
\begin{gather}\label{eqcont}
 G \pfsum H \neq \Hi(\W)
\end{gather}
and assume if possible, that there exists
a constructed stream $\omega^{j_\star}$ which compresses to white noise. Although, we cannot assume that any $H^j$ is state synchronized with $G$ directly, we can consider
the structure of both $G$ and $H$ to be represented (without loss of generality) by the one for $G\times H$, in which they can be assumed to be synchronized (since state $q_i$ in G and $q_k$ in H can
be mapped to state $(q_i,q_k)$ in $G \times H$). Denoting the machines modified by the synchronous product as $G^\times$ and $H^\times$ respectively, we note:
\begin{gather}
 G^\times + H^\times = \Hi(\W) \ \mathrm{(By \ Assumption)}
\end{gather}
But since $\HA(G^\times)= \HA(G)$ and $\HA(H^\times)= \HA(H)$ (since the underlying measures are not modified by going to a non-minimal realization via synchronous product), it follows:
\begin{gather}
 \HA(G) + \HA(H) = \W \Rightarrow G \pfsum H = \Hi(\W)
\end{gather}
which contradicts Eq.~\eqref{eqcont}. This completes the proof.
\end{IEEEproof}

Our key motivation for developing the annihilator was to be able to classify PFSA-based patterns faster and in a more robust fashion in real-time or near-real-time field operation.
The argument for robustness is pretty obvious, since  one state models, especially with uniform generation probabilities of the symbols ($i.e.$ white noise) are the easiest ones to identify
reliably  for any compression algorithm. The argument for fast identification is more involved, primarily due to the fact that the annihilators selectively erase symbols leading to a decrease
in the lengths of the observed symbol strings. Thus, although we only need to check for white noise in the outputs (which is significantly faster compared to directly identifying the original pattern),
the fact that now we are dealing with a shorter string, implies that there is the possibility that the increased speed of identification is offset by the slow down of the rate of symbol production at the outputs. In the next section, we investigate this issue in more details, and derive rigorous performance guarantees.
\vspace{0pt}
\section{Performance Of Semantic Annihilators}\label{secperf}
We need the notion of a stationary distribution on the states of a given PFSA. This
is in fact the stationary distribution for the stochastic transition probability matrix that can be computed
from the connectivity graph and the symbol generation probabilities $\widetilde{\Pi}$.
Also, as stated before, we assume that all PFSA considered in this paper are irreducible, $i.e.$, have a strongly connected graph and hence yields an irreducible
transition probability matrix.
\begin{defn}[Stationary Distribution] For a given PFSA $G=(Q,\Sigma,\delta, \widetilde{\Pi})$, the stationary distribution $\wp^G \in [0,1]^{\vert Q\vert},$ $ \sum_i \wp^G_i = 1$ is defined as:
\begin{enumerate}
\item Construct the transition probability matrix $\Pi$ as:
\begin{gather}
 \forall q_i,q_j \in Q,\ \Pi\big\vert_{ij} = \sum_{\sigma_k: \delta(q_i,\sigma) = q_j}\widetilde{\Pi}(q_i,\sigma)
\end{gather}
\item Noting that $\Pi$ is an irreducible  stochastic matrix, compute the stationary distribution $\wp^G$ as the stationary probability distribution for the
state transition matrix $\Pi$, $i.e.$, $\wp^G$ is the unique  sum-normalized left eigenvector for $\widetilde{\Pi}$ satisfying
 $\wp^G\Pi = \wp^G$.

%
\end{enumerate}
 \end{defn}
It follows from the irreducibility assumption, that the stationary distribution is unique for a given PFSA, and has no dependence on the
initial state~\cite{BP94}.
\begin{notn}
In the sequel, we use the notation: $\wp^G_\star = \min_{q_i \in Q} \wp^G$.
\end{notn}
Also, our assumption of irreducible models leads to the following property for the stationary distribution:
\begin{prop}\label{propnonz}
 For any PFSA $G=(Q,\Sigma,\delta, \widetilde{\Pi})$ with an irreducible underlying graph, $\wp^G_\star > 0$.
\end{prop}
\vspace{4pt}\begin{IEEEproof}
 Since $\Pi$ is irreducible for such $G$, no non-negative left eigenvector of $\Pi$ has a zero coordinate~\cite{BP94}.
\end{IEEEproof}
We want to estimate the shortening experienced by the sensed symbol strings due to the annihilation operation. We require the notion
of the auxiliary PFSA $\aux(G)$
for a given PFSA $G$, which captures the
simultaneous operation of the two machines, without erasure of the
non-matching symbols.
\begin{defn}[Auxiliary PFSA]\label{defaux}
For a given PFSA $G=(Q,\Sigma,\delta, \widetilde{\Pi})$, the auxiliary PFSA $\aux(G)$ is defined as:
$\aux(G) = (Q,\Sigma\bigcup\Sigma',\delta^\aux, \widetilde{\Pi}^\aux)$, where $\Sigma'$ is a \textit{distinct} isomorphic copy of $ \Sigma$, with
$\mathscr{I}:\Sigma \rightarrow \Sigma'$ being the (bijective) isomorphism, and:
\begin{subequations}
\begin{gather}
 \delta^\aux(q_i,\sigma) = \left \{ \begin{array}{ll}
                            \delta(q_i,\sigma) & \mathrm{if} \ \sigma \in \Sigma \\
\delta(q_i,\mathscr{I}^{-1}\sigma) & \mathrm{otherwise}
                                           \end{array}
\right.\\
\widetilde{\Pi}^\aux(q_i,\sigma) = \left \{\begin{array}{ll}
                                                   \frac{1}{\vert \Sigma \vert}\mathscr{H}_i & \mathrm{if} \ \sigma \in \Sigma \\
                        \widetilde{\Pi}(q_i,\sigma) - \frac{1}{\vert \Sigma \vert}\mathscr{H}_i & \mathrm{otherwise}
                                                   \end{array}
\right.\label{eqaux}
\end{gather}
\end{subequations}
where $\mathscr{H}_i$ is the harmonic mean of the $i^{th}$ row of the $\widetilde{\Pi}$ matrix for $G$.
\end{defn}
\begin{prop}[Properties of the Auxiliary Automaton]\label{propauxilliary}
The auxiliary automaton $\aux(G) = (Q,\Sigma\bigcup\Sigma',\delta^\aux, \widetilde{\Pi}^\aux)$ has the following properties:
\begin{enumerate}
 \item $\wp^{\aux(G)} =\wp^G$
\item If $H$ is the annihilator component that is correctly state-synchronized with $G$ (where $G$ is the correct PFSA corresponding to the annihilator), then $\aux(G)$ correctly tracks $H$ (state-wise and symbol-wise), if we consider that all $\sigma\in \Sigma'$ are unobservable.
\end{enumerate}
\end{prop}
\vspace{4pt}\begin{IEEEproof}
 $(1)$ follows immediately from Definition~\ref{defaux}, by noting that the probability transition matrix is left unaltered in the construction of $\aux(G)$.
For $(2)$, we  note that the transition structure for $H$ (and hence $G$) is recovered if we map $\forall \sigma \in \Sigma', \ \sigma \mapsto \mathscr{I}^{-1}\sigma$.
Next, we compute the probability $p_{obs}(q_i,\sigma)$ of an observable $\sigma$ when $H$ is at state $q_i$ as:
\begin{align}\label{eqcont111}
&\forall \sigma \in \Sigma, \notag\\ p_{obs}(q_i,\sigma) =& \widetilde{\Pi}^G(q_i,\sigma)\widetilde{\Pi}^H(q_i,\sigma)\notag \\
=& \widetilde{\Pi}^G(q_i,\sigma)(\widetilde{\Pi}^G(q_i,\sigma))^{-1}\frac{1}{\sum_\sigma (\widetilde{\Pi}^G(q_i,\sigma))^{-1}}\notag \\ =& \frac{1}{\vert\Sigma\vert} \mathscr{H}_i
\end{align}
It follows from above, that the probability of an unobservable $\sigma$ when $H$ is at state $q_i$ is given by:
\begin{gather}\label{equnobs}
\forall \sigma \in \Sigma', p_{unobs}(q_i,\sigma) = \widetilde{\Pi}^G(q_i,\sigma) - \frac{1}{\vert \Sigma \vert}\mathscr{H}_i
\end{gather}
which completes the proof.
\end{IEEEproof}
\begin{cor}\label{coraux}
(To Proposition~\ref{propauxilliary}) If the length of the  symbol string generated by $G$  is denoted by $L_G$, then the  length $L_{ann}$ of the correctly
annihilated string  satisfies:
\begin{gather}
 \lim_{L_G \rightarrow \infty}\frac{L_{ann}}{L_G}= \sum_i^{\vert Q \vert} \wp^G_i\mathscr{H}_i
\end{gather}

\end{cor}
\vspace{4pt}\begin{IEEEproof}
We first note that the stationary frequency distribution $\vartheta_\Sigma$ of the symbols (over alphabet $\Sigma$) in a string generated by an arbitrary  irreducible PFSA $G_{arb}$ is given by:
\begin{gather}
 \vartheta_\Sigma = \wp^{G_{arb}}\widetilde{\Pi}^{G_{arb}}
\end{gather}
where the independence from the initial state follows from the irreducibility of $G_{arb}$. It then follows from Proposition~\ref{propauxilliary}, that the
frequency distribution for the auxiliary automaton $\aux(G)$ is given by:
\begin{gather}
 \vartheta_{\Sigma \bigcup \Sigma'} = \wp^{\aux(G)}\widetilde{\Pi}^{\aux(G)} = \wp^G\widetilde{\Pi}^{\aux(G)}
\end{gather}
which in turn implies (See Eq.~\eqref{eqaux}) that the probability $\lambda$ that any symbol generated by $G$ is observable is given by:
\begin{gather}
 \lambda = \sum_{i:\sigma_i \in \Sigma}\wp^G\widetilde{\Pi}^{\aux(G)}\big \vert_i = \sum_i^{\vert Q \vert} \wp^G_i\mathscr{H}_i
\end{gather}
This completes the proof.
\end{IEEEproof}
Next, we define the coefficient of annihilation advantage:
\begin{defn}[Coefficient of Annihilation Advantage]\label{defbeta} For a given PFSA $G=(Q,\Sigma,\delta,\widetilde{\Pi})$, let $L_d$ be the string length required for
direct identification via semantic compression, and let $L_w$ be the string length required for identifying
symbolic white noise. Then the Coefficient of Annihilation Advantage ($\beta$) is defined as the ratio:
\begin{gather}
 \beta = \frac{L_w}{L_d \sum_i^{\vert Q \vert} \wp^G_i\mathscr{H}_i}
\end{gather}
\end{defn}
\begin{rem}
It follows that when we have enough
data to do a direct compression (of say length $L_d$), then the expected length of the correctly annihilated string is given by $L_d \sum_i^{\vert Q \vert} \wp^G_i\mathscr{H}_i$
Since we are required to identify symbolic white noise at the annihilator output, and if the string length for identification of symbolic white noise is denoted by $L_w$,
then identification via semantic annihilation is advantageous if
 we have $L_w < L_d\sum_i^{\vert Q \vert} \wp^G_i\mathscr{H}_i$, $i.e.$, if we have $\beta < 1$.
\end{rem}

In the sequel, we compute upper bounds on the Coefficient of Annihilation Advantage $\beta$.
In order to do so, it is obvious that we need to relate the lengths $L_d$ and $L_w$. However, we wish to achieve this without
reference to any specific algorithm for semantic compression, $i.e.$, we want the computed bounds to hold true irrespective of the manner we construct PFSA models out of symbol strings.
We note that if we  are to compress a string from a symbolic white noise, then we would expect to obtain a single state PFSA with equi-probable symbols. However, since we are talking about
probabilistic generators, observing $1$ symbol each from the
alphabet would not be sufficient; or rather would be a very bad way of inferring that the symbol string is generated from the symbolic white noise. Since we assume that $L_w$ is the string length required for the identification (for the particular algorithm, whichever that may be), the number of symbols of each label that we need to observe would be at least $\frac{1}{\vert \Sigma\vert }L_w$. In the sequel, we assume that for an arbitrary PFSA,
the number of symbols of each label that we need to observe at each state must also  be of at least this value $\frac{1}{\vert \Sigma\vert}L_w$, since the chosen algorithm apparently requires this many observations
for statistical inference.

\begin{prop}[Upper Bound for $\beta$]\label{propbounds}
For a given PFSA $G=(Q,\Sigma,\delta,\widetilde{\Pi})$ with an irreducible underlying graph, which is not a realization of symbolic white noise, we have the following upper bounds:
\begin{subequations}
 \begin{align}
1)\mspace{50mu}  &\beta = \frac{L_w}{L_d \sum_i^{\vert Q \vert} \wp^G_i\mathscr{H}_i} < \frac{\vert\Sigma\vert\vert Q\vert\wp^G_\star}{\displaystyle\sum_{i=1}^{\vert Q \vert} \wp^G_i\mathscr{H}_i \sum_{j=1}^{\vert Q \vert} \mathscr{H}_j^{-1}} \triangleq \beta_1\\
2)\mspace{50mu} &\beta_1 \leqq \frac{\vert\Sigma\vert}{\vert Q\vert}
 \end{align}
\end{subequations}
\end{prop}
\vspace{4pt}\begin{IEEEproof}
We note for each state $q_i\in Q$, we have:
\begin{gather}
 \wp^G_\star L_d \min_{\sigma \in \Sigma}\widetilde{\Pi}(q_i,\sigma) \geqq \frac{1}{\vert \Sigma\vert}L_w
\end{gather}
which  follows from noting that $\wp^G_\star L_d$ is at least the number of times state $q_i$ is visited, and hence $\wp^G_\star L_d \min_{\sigma \in \Sigma}\widetilde{\Pi}(q_i,\sigma)$
is at least the expected number of the least likely symbols generated at $q_i$. It follows:
\begin{align}
 L_d &\geqq \frac{L_w}{\vert \Sigma\vert \wp^G_\star} \frac{1}{\min_{\sigma \in \Sigma}\widetilde{\Pi}(q_i,\sigma)} \geqq \frac{L_w}{\vert \Sigma\vert \wp^G_\star} \mathscr{H}_i^{-1} \\
&\Rightarrow \vert Q\vert L_d > \frac{L_w}{\vert \Sigma\vert \wp^G_\star} \sum_{i=1}^{\vert Q \vert}\mathscr{H}_i^{-1} \label{eqstr}\\
&\Rightarrow \beta = \frac{L_w}{L_d \sum_i^{\vert Q \vert} \wp^G_i\mathscr{H}_i} < \frac{\vert\Sigma\vert\vert Q\vert\wp^G_\star}{\displaystyle\sum_{i=1}^{\vert Q \vert} \wp^G_i\mathscr{H}_i \sum_{j=1}^{\vert Q \vert} \mathscr{H}_j^{-1}}
\end{align}
Note the strict bound in the second inequality in Eq.~\eqref{eqstr} follows from the fact that $G$ is not a realization of symbolic white noise, implying $\exists q_i \in Q, \ \mathscr{H}_i > \min_{\sigma \in \Sigma}\widetilde{\Pi}(q_i,\sigma)$,
which completes the proof of Statement $(1)$. For Statement $(2)$, we first note that for any sequence of real numbers, the harmonic mean of the sequence is bounded above by its arithmetic mean. Hence, it follows that:
\begin{align*}
 &\frac{\displaystyle\vert Q \vert }{\displaystyle\sum_{j=1}^{\vert Q \vert} \mathscr{H}_j^{-1}} \leqq \frac{\displaystyle\sum_{j=1}^{\vert Q \vert} \mathscr{H}_j}{\vert Q \vert } \Rightarrow \frac{1}{\displaystyle\sum_{j=1}^{\vert Q \vert} \mathscr{H}_j\sum_{j=1}^{\vert Q \vert} \mathscr{H}_j^{-1} } \leqq \frac{1}{\vert Q \vert^2} \notag \\
\Rightarrow & \frac{1}{\displaystyle\sum_{j=1}^{\vert Q \vert} \wp^G_j\mathscr{H}_j\sum_{j=1}^{\vert Q \vert} \mathscr{H}_j^{-1} } \leqq \frac{1}{\displaystyle\sum_{j=1}^{\vert Q \vert} \wp^G_\star\mathscr{H}_j\sum_{j=1}^{\vert Q \vert} \mathscr{H}_j^{-1} } \leqq \frac{1}{\wp^G_\star\vert Q \vert^2} \notag \\
\Rightarrow & \frac{\vert\Sigma\vert\vert Q\vert\wp^G_\star}{\displaystyle\sum_{i=1}^{\vert Q \vert} \wp^G_i\mathscr{H}_i \sum_{j=1}^{\vert Q \vert} \mathscr{H}_j^{-1}} \leqq \frac{\vert \Sigma \vert \vert Q \vert \wp^G_\star}{\vert Q \vert^2 \wp^G_\star} = \frac{\vert\Sigma\vert}{\vert Q\vert}
\end{align*}
where the last step follows from the fact that irreducibility of $G$ guarantees $\wp^G_\star > 0$. This completes the proof.
\end{IEEEproof}
\begin{rem}
 Note that although we assume that $G$ is not a realization of symbolic white noise, we could not assume $\exists q_i \in Q \ \wp^G_i > \wp^G_\star$, which would have made the
bound in Statement $(2)$ strict. The reason is that it is possible for a PFSA to have non-uniform symbol generation probabilities from some states, and yet end up having an uniform stationary distribution over its states.
Note here that the property of being white (in the way we defined) has to do with the uniformity of the rows of the $\widetilde{\Pi}$ matrix, and not the stationary probabilities.
\end{rem}

\begin{rem}
Proposition~\ref{propbounds} is a strong result which implies that pattern classification via semantic annihilators is in fact advantageous for most PFSA encountered in practice, where typically
one has a relatively small number of alphabet symbols and a possibly large number of machine states.
\end{rem}
\begin{rem}
 The bounds computed in Proposition~\ref{propbounds} are not tight. Specifically, note that we neglected the fact that for a general PFSA, the string length for identification could be
significantly  greater due to issues relating to adequately achieving statistical stationarity of the observed stream. Thus even for models for which $\vert\Sigma\vert > \vert Q \vert$, it is not automatic
that identification via annihilation is slower compared to direct compression.
\end{rem}
\vspace{0pt}
\section{Summarized Algorithms for Classification Via Semantic Annihilation}\label{secalgo}
For each pattern in the specified pattern library, we first compute the inverse PFSA using Algorithm~\ref{Algorithm03}. Note, that
step 4 in Algorithm~\ref{Algorithm03} is well-defined (and does not encounter a divide-by-zero overflow) on account of our
assumption of the restricted set $\mathscr{A}^+$ (See Definition~\ref{defnA+}). Once the inverse patterns are computed, we need to set up
the pattern-specific annihilators. Namely, for each pattern with $\vert Q\vert$ states, we need $\vert Q\vert$ copies of the inverse, each initialized
to a distinct state, as stated before. The annihilation process requires sequential generation of symbols from these initialized PFSA, in accordance
to their computed morph matrices. This is done as follows:
\begin{enumerate}
\item Given the current state, we first select the corresponding row of the morph matrix, which
specifies the probability distribution of the to-be-generated symbol over the alphabet.
\item We generate a symbol in accordance to this distribution.
\end{enumerate}
There are standard reported ways of selecting an outcome in accordance to a specified distribution. We explicitly state one method involving
a uniform random number generator with  range $[0,1]$, which
guarantees that the asymptotic time-complexity of this choice is $O(\log_2(\vert \Sigma \vert) )$ (See Algorithm~\ref{Algorithm01}). The stated approach involves
considering the cumulative distribution for the symbol. Since this has to be done each time a symbol is generated, we compute the
cumulative morph matrix $\widetilde{\Pi}_{cum}$ for the inverted models offline as follows:
\begin{defn}[Cumulative Morph Matrix]
 The cumulative morph matrix $\widetilde{\Pi}_{cum}$ is computed as follows:
\begin{gather}
 \widetilde{\Pi}_{cum} \bigg \vert_{ij} = \sum_{r=1}^j \widetilde{\Pi} \bigg \vert_{ir}
\end{gather}

\end{defn}
The sequential symbol generation then uses rows of the cumulative morph matrix instead, as the input $\nu$ to Algorithm~\ref{Algorithm01}.
\begin{lem}
 Assuming that uniform random numbers in the range $[0,1]$ can be generated in constant time, the asymptotic time-complexity of Algorithm~\ref{Algorithm01} is $O(\log_2(\vert \Sigma \vert) )$.
\end{lem}
\begin{IEEEproof}
 We note that the possible number of choices for the to-be-generated symbol reduces by half its previous value in each iteration, implying that
the number of iterations $I$ satisfies:
\begin{gather*}
 2^I \leqq \vert \Sigma \vert
\Rightarrow I \leqq \log_2(\vert \Sigma \vert)
\end{gather*}
which completes the proof.
\end{IEEEproof}
Each copy of the inverted model in the annihilator accesses the sensed symbol, generates its own symbol
in accordance to its current state, reports the symbol if there is a match, and finally updates the current state
using the sensed symbol. The sequence of moves for each component (or copy) is enumerated in Algorithm~\ref{Algorithm02}. The
$\vert Q\vert$ reported streams are individually compressed to check if any is in fact white noise.

\begin{algorithm}[t]
 \SetLine
  \SetKwData{Left}{left}
  \SetKwData{This}{this}
  \SetKwData{Up}{up}
  \SetKwFunction{Union}{Union}
  \SetKwFunction{FindCompress}{FindCompress}
  \SetKwInOut{Input}{input}
  \SetKwInOut{Output}{output}
  \SetKw{low}{lowerB}
  \SetKw{up}{upperB}
 \SetKw{Tr}{true}
   \SetKw{Tf}{false}
  \caption{Computation of Inverse Pattern}\label{Algorithm03}
\Input{PFSA $G=(Q,\Sigma,\delta,\widetilde{\Pi})$}
\Output{PFSA $-G$}
\Begin{
\For{$i=1:\vert Q\vert$}{
\For{$j=1:\vert \Sigma\vert$}{
$\displaystyle\widetilde{\Pi}'_{ij} = \frac{1}{\widetilde{\Pi}_{ij}}$\;
}
\For{$j=1:\vert \Sigma\vert$}{
$\displaystyle\widetilde{\Pi}'_{ij} = \frac{\widetilde{\Pi}'_{ij}}{\sum_j\widetilde{\Pi}'_{ij}}$\;
}

}
$-G=(Q,\Sigma,\delta,\widetilde{\Pi}')$\;
}
\end{algorithm}

\begin{algorithm}[!ht]
\SetLine
\linesnumbered
   \SetKwInOut{Input}{input}
  \SetKwInOut{Output}{output}
  \SetKw{low}{lowerB}
  \SetKw{up}{upperB}
 \SetKw{Tr}{true}
   \SetKw{Tf}{false}
  \caption{Probabilistic Symbol Generation}\label{Algorithm01}
\Input{Non-negative vector $\nu$ of length $\vert \Sigma\vert$ such that $\nu\vert_i \leqq \nu\vert_{i+1}, \nu\vert_{\vert\Sigma\vert}=1$,
Alphabet $\Sigma= \{\sigma_1,\cdots,\sigma_{\vert \Sigma \vert}\}$}
\Output{Generated Symbol $\sigma$}
\Begin{
Generate random key $K_r \in [0,1]$ \;
\low = 1\;
\up = $\vert \Sigma \vert$ \;
\While{ $\textrm{\bf upperB} >  (\textrm{\bf lowerB}+ 1)$ }{
M = $\lceil \frac{\textbf{upperB} - \textbf{lowerB}}{2} \rceil$\tcc*[r]{Rounding to Next Integer}
\eIf{ $K_r \leqq \nu \vert_M$}{\up = M\;
}{\low = M\;
}
}
\eIf{$k_r \leqq \nu \vert_{\textbf{lowerB}} $}
{\vspace{3pt}$\sigma = \sigma_{\textbf{lowerB}}$\;}
{$\sigma = \sigma_{\textbf{upperB}}$\;}
}
\end{algorithm}
%
\begin{algorithm}[!ht]
 \SetLine
  \SetKwData{Left}{left}
  \SetKwData{This}{this}
  \SetKwData{Up}{up}
  \SetKwFunction{Union}{Union}
  \SetKwFunction{FindCompress}{FindCompress}
  \SetKwInOut{Input}{input}
  \SetKwInOut{Output}{output}
  \SetKw{low}{lowerB}
  \SetKw{up}{upperB}
 \SetKw{Tr}{true}
   \SetKw{Tf}{false}
  \caption{Componentwise  Annihilation Operation}\label{Algorithm02}
\Input{Cumulative Morph Matrix $\widetilde{\Pi}_{cum}$, Initial state $q_{\textrm{\bf init}}$, Transition function $\delta$}
\Output{Reported symbol stream}
\Begin{
Set current state $q_{\textrm{\bf curr}} = q_{\textrm{\bf init}}$\;
\tcc{Infinite loop}
\While{\Tr}{
Observe sensed symbol $\sigma_{\textrm{\bf sensed}}$ \;
Generate random symbol $\sigma_{\textrm{\bf gen}}$ using  row of $\widetilde{\Pi}_{cum}$ corresponding to $q_{\textrm{\bf curr}}$\tcc*[r]{Algorithm~\ref{Algorithm01}}
\If{$\sigma_{\textrm{\bf sensed}} == \sigma_{\textrm{\bf gen}}$}
{ \Darkred Report  $\sigma_{\textrm{\bf gen}}$\tcc*[r]{Annihilated Stream} }
Update current state $q_{\textrm{\bf curr}} = \delta(q_{\textrm{\bf curr}},\sigma_{\textrm{\bf sensed}})$\;
}
}
\end{algorithm}

\section{Asymptotic Complexity Analysis}\label{seccomplex}
We ascertain the asymptotic time complexity per sensed symbol of the online portion of the annihilation process, assuming
the pattern corresponding to the annihilator is indeed present in the sensed stream. This analysis is important since
the annihilator is processing a multi-stream input, and we need to convince ourselves that the work required
per observed symbol is not too great, particularly since an overtly complex algorithm will be unable to handle high
data rates.

We assume, as before, that random keys can be generated in constant time. Then, we have the following result:
\begin{prop}\label{propComplexity}
For a given PFSA $G=(Q,\Sigma,\delta,\widetilde{\Pi})$, the asymptotic time-complexity $\mathscr{A}_G$ of classification via annihilation, per sensed symbol, is bounded as:
\begin{gather}
 \mathscr{A}_G = O(\log_2(\vert \Sigma \vert))
\end{gather}
\end{prop}
\begin{IEEEproof}
Time-complexity of identification $C_1$, considering all $\vert Q\vert$ components of the annihilator, satisfies:
\begin{gather}
  C_1 \leqq T_R \vert Q\vert O(\log_2(\vert \Sigma \vert)) L_w  C_0
\end{gather}
where $T_R$ is the complexity of generating random keys in the range $[0,1]$, $L_w$ is the string length required for
identifying the symbolic white noise, and $C_0$ is the time complexity of identifying symbolic white noise (using some given direct compression algorithm, and assuming we
check for white noise on each stream after each sensed symbol observation).
Hence, assuming that we have the total sensed string length as $L_d$, it follows that the time complexity per sensed symbol is bounded by:
\begin{subequations}
\begin{align}
 & \mathscr{A}_G\leqq T_R C_0\vert Q\vert O(\log_2(\vert \Sigma \vert)) \frac{L_w}{L_d} \label{eqa}
 \intertext{Using Definition~\ref{defbeta}, we have: }
 &\mathscr{A}_G \leqq T_R C_0 \vert Q\vert O(\log_2(\vert \Sigma \vert)) \beta \sum_i^{\vert Q \vert}  \wp^G_i\mathscr{H}_i 
 \intertext{Using Proposition~\ref{propbounds}, and noting $\mathscr{H}_i \leqq \frac{1}{\vert \Sigma \vert}$, we have:}
  &\mathscr{A}_G \leqq T_R C_0 \vert Q\vert O(\log_2(\vert \Sigma \vert)) \frac{\vert \Sigma \vert}{\vert Q\vert} \sum_i^{\vert Q \vert} \wp^G_i\frac{1}{\vert \Sigma \vert}  \\
\Rightarrow &\mathscr{A}_G \leqq T_R C_0 O(\log_2(\vert \Sigma \vert))
\intertext{Neglecting constant time factors, we have:}
 &\mathscr{A}_G = O(\log_2(\vert \Sigma \vert))
\end{align}
\end{subequations}
Note that Eqn.~\eqref{eqa} is exact and not an averaging, since we do the same work every time a symbol is sensed. This
completes the proof.
\end{IEEEproof}
This is a strong result showing  that the  asymptotic time-complexity  of classification via annihilation, per symbol, is independent of complexity of the
pattern and the number of PFSA states, and is only mildly dependent on the cardinality of the alphabet. Again, since the alphabet sizes are
relatively small, and recalling that the proposed technique is provably faster compared to direct compression for most models, it follows that
classification via annihilation is indeed  highly advantageous for online operation.
\section{Verification \& Validation}\label{secvv}
\begin{figure}[t]
\centering
\psfrag{121}[cb]{\Darkred\scriptsize\txt{$\phantom{x}$ {\bf \Nblue 1} $\vert$ 0.3}}
\psfrag{022}[cb]{\Darkred\scriptsize\txt{{\bf \Nblue 0} $\vert$ 0.7 }}
\psfrag{011}[cb]{\Darkred\scriptsize\txt{{\bf \Nblue 0} $\vert$ 0.2}}
\psfrag{031}[cb]{\Darkred\scriptsize\txt{{\bf \Nblue 0} $\vert$ 0.9 }}
\psfrag{112}[cb]{\Darkred\scriptsize\txt{{\bf \Nblue 1} $\vert$ 0.8 }}
\psfrag{124}[cb]{\Darkred\scriptsize\txt{{\bf \Nblue 1} $\vert$ 0.9 }}
\psfrag{132}[cb]{\Darkred\scriptsize\txt{{\bf \Nblue 1} $\vert$ 0.1 }}
\psfrag{043}[bc]{\Darkred\scriptsize\txt{$\phantom{xX}${\bf \Nblue 0} $\vert$ 0.7 }}
\psfrag{144}[cb]{\Darkred\scriptsize\txt{$\phantom{x}${\bf \Nblue 1} $\vert$ 0.3 }}
\psfrag{023}[cb]{\Darkred\scriptsize\txt{{\bf \Nblue 0} $\vert$ 0.1}}
 \psfrag{q1}[lc]{$\boldsymbol{q_1}$}
\psfrag{q2}[lc]{$\boldsymbol{q_2}$}
\psfrag{q3}[lc]{$\boldsymbol{q_3}$}
\psfrag{q4}[lc]{$\boldsymbol{q_4}$}
\subfloat[M2]{\includegraphics[width=2.5in]{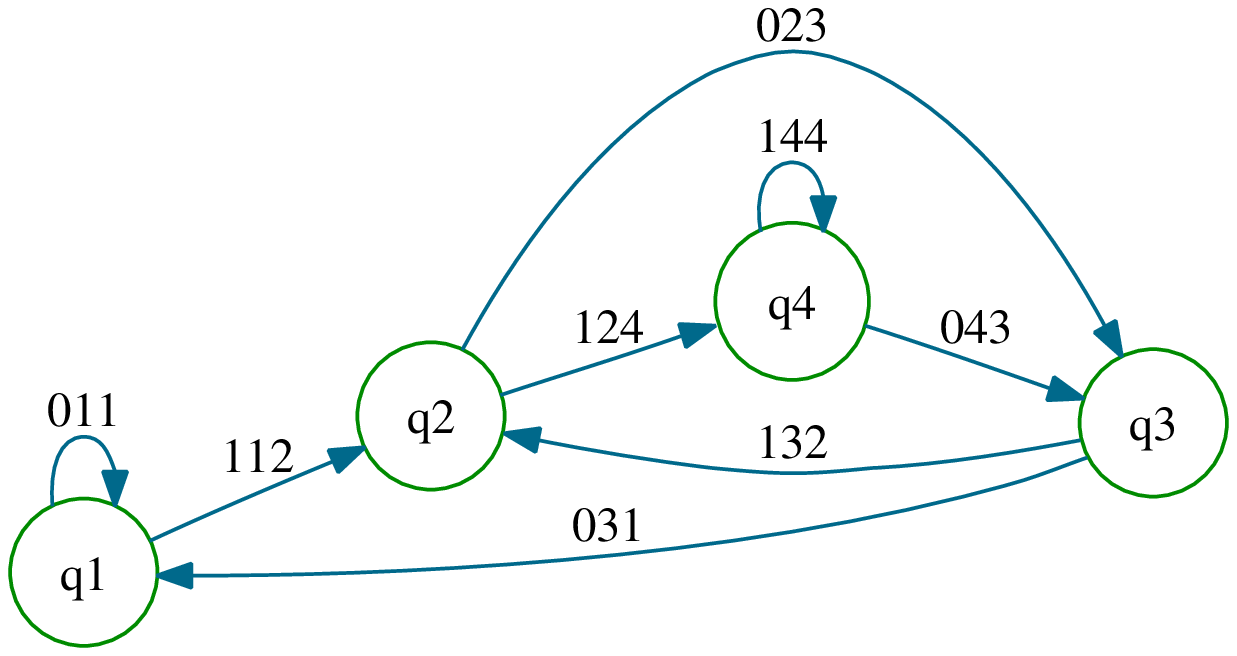}}\hspace{40pt}
 \psfrag{q1}[lb]{$\boldsymbol{q_1}$}
\psfrag{q2}[lb]{$\boldsymbol{q_2}$}
\subfloat[S1]{\includegraphics[width=1.5in]{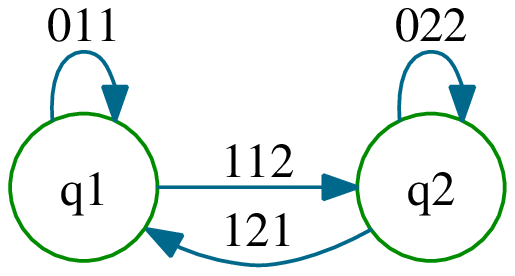}}
\caption{PFSA models used for simulation, real numbers are symbol probabilities, integers are symbol labels: (a) M2 is a subshift of finite type having the structure of a
suffix automaton, (b)S1 is a generalized even shift process, is a strictly Sofic system and has no synchronizing string}\label{figsimmodel1}
\end{figure}
In this section, we validate the preceding theoretical developments in simulation.
The PFSA models selected for generating the simulated symbol string is illustrated in Figure~\ref{figsimmodel1}.
%
The model (M2) shown in Figure~\ref{figsimmodel1}(a)
has the structure of a suffix automaton~\cite{Mu96}. PFSA which have such structures are easier to identify from
symbolic strings; primarily due to the existence of synchronizing strings~\cite{SS04} \textit{(strings which lead to a particular state irrespective of the starting state)}. For example, in
the model M2, the states $q_1,q_2,q_3,q_4$ can be easily seen to represent sets of symbol strings ending in
$00, 01, 10, 11$ respectively~\cite{R04,Mu96}.
Although, the state structure is not available a priori to the compression algorithm, nevertheless, such
so-called $d$-Markov machines~\cite{R04} are significantly easier to identify. For examples of physical situations in anomaly detection which give rise to, or are effectively modeled by such  $d$-Markov machines, the reader is referred to \cite{CRR05,RR06}. The second model (S1) (Figure~\ref{figsimmodel1}(b)) has only two states. However,
S1 represents a generalization of the even-shift process, and its underlying graph is an example of
a strictly Sofic shift process \textit{(and not a sub-shift of finite type~\cite{LM95})}. Specifically, S1 does not have any synchronizing strings, $i.e.$, without the knowledge of the initial state one cannot infer the current state in a deterministic sense even from arbitrary long observation strings. Such models are significantly more difficult to identify (See \cite{SS04} for discussion) for any of the compression algorithms reported in the literature~\cite{Mu96,R04}. 
\begin{figure*}[!t]
\centering
\includegraphics[width=6.5in]{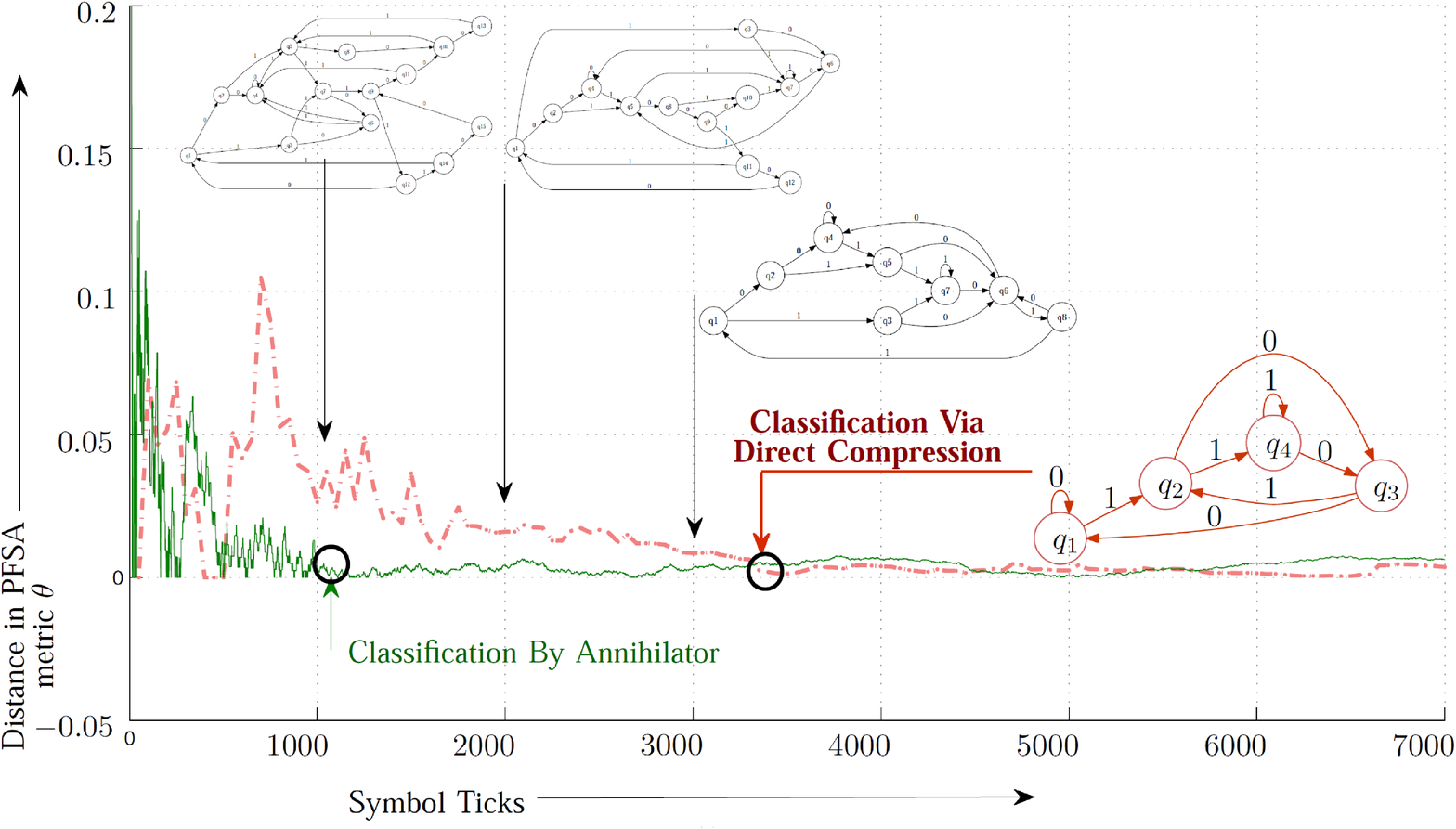}
\vspace{0pt}
\centering
\psfrag{1000}[tc]{$1000$}
\psfrag{2000}[tc]{$2000$}
\psfrag{3000}[tc]{$3000$}
\psfrag{4000}[tc]{$4000$}
\psfrag{5000}[tc]{$5000$}
\psfrag{6000}[tc]{$6000$}
\psfrag{7000}[tc]{$7000$}
\psfrag{0}[cc]{$0$}
\psfrag{0.1}[cc]{$0.1\phantom{x}$}
\psfrag{0.2}[cc]{$0.2\phantom{x}$}
\psfrag{0.3}[cc]{$0.3\phantom{x}$}
\psfrag{0.4}[cc]{$0.4\phantom{x}$}
\psfrag{0.5}[cc]{$0.5\phantom{x}$}
\psfrag{0.6}[cc]{$0.6\phantom{x}$}
\psfrag{0.7}[cc]{$0.7\phantom{x}$}
\psfrag{0.8}[cc]{$0.8\phantom{x}$}
\psfrag{                                   A}[cc]{}
\psfrag{C}[lc]{\scriptsize \Mblue Annihilator}
\psfrag{B}[lc]{\scriptsize \Mblue \txt{$\phantom{o}$\\Direct Cmp.}}
\psfrag{P}[cr]{\txt{Metric $\theta$}}
\psfrag{D}[cr]{\txt{$\phantom{x}$\\Symbol Ticks}}
\psfrag{I}[bl]{\txt{\Dgreen $\phantom{x}$Classification \\ \Dgreen By Annihilator}}
\subfloat[]{\includegraphics[width=3.1in]{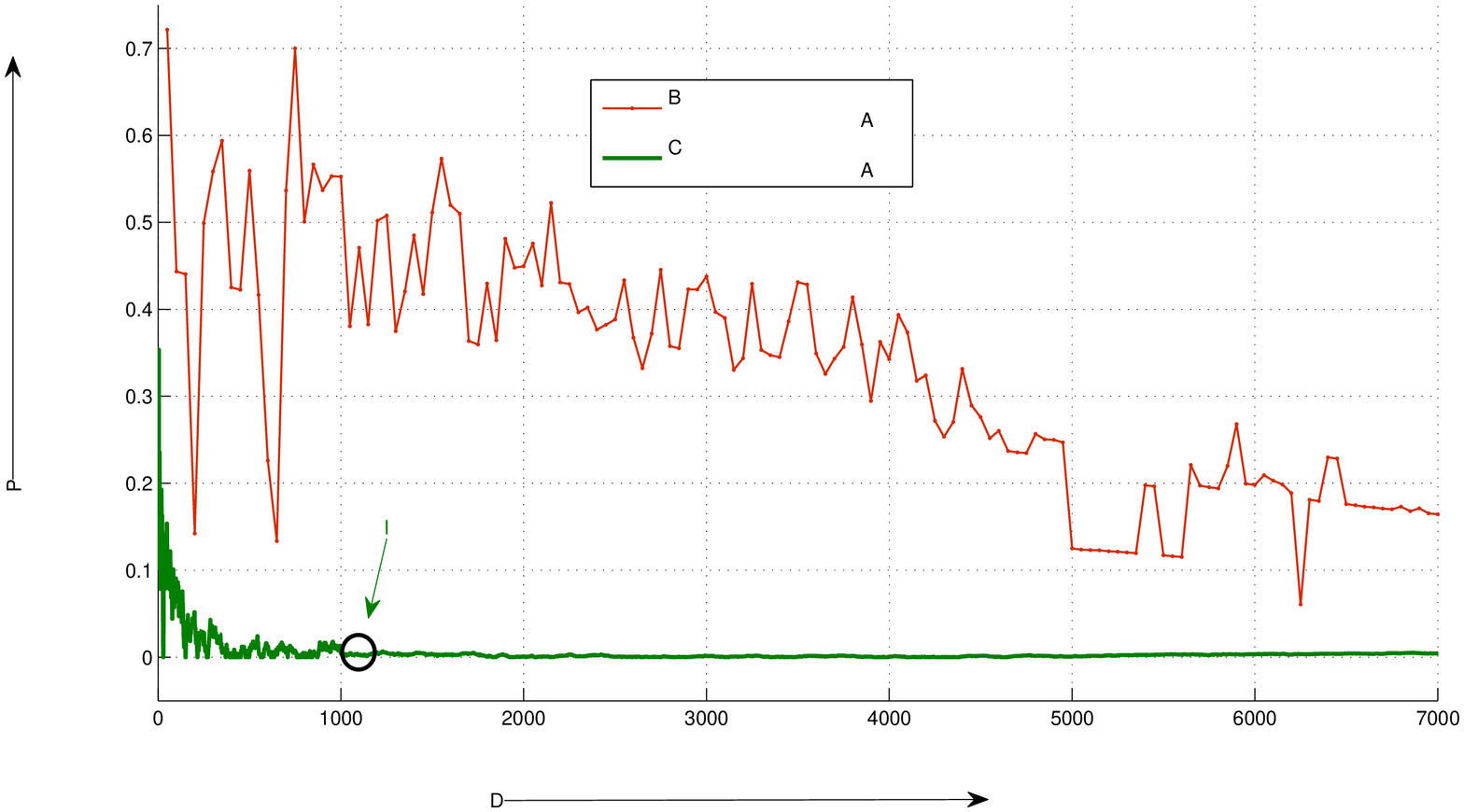}}
\hspace{40pt}
\psfrag{1000}[tc]{$1000$}
\psfrag{500}[tc]{$500$}
\psfrag{2000}[tc]{$2000$}
\psfrag{1500}[tc]{$1500$}
\psfrag{2500}[tc]{$2500$}
\psfrag{6000}[tc]{$6000$}
\psfrag{7000}[tc]{$7000$}
\psfrag{0}[cc]{}
\psfrag{0.1}[cc]{$0.1\phantom{x}$}
\psfrag{0.2}[cc]{$0.2\phantom{x}$}
\psfrag{0.3}[cc]{$0.3\phantom{x}$}
\psfrag{0.4}[cc]{$0.4\phantom{x}$}
\psfrag{0.5}[cc]{$0.5\phantom{x}$}
\psfrag{0.6}[cc]{$0.6\phantom{x}$}
\psfrag{0.7}[cc]{$0.7\phantom{x}$}
\psfrag{0.8}[cc]{$0.8\phantom{x}$}
\psfrag{                                                 A}[cc]{}
\psfrag{C}[lc]{\scriptsize \Mblue Annihilator}
\psfrag{B}[lc]{}
\psfrag{m}[lc]{\scriptsize \Mblue \txt{$\phantom{.}$\\Pattern Match}}
\psfrag{nm}[lc]{\scriptsize \Mblue \txt{$\phantom{.}$\\No Match}}
\psfrag{P}[cr]{\txt{Metric $\theta$}}
\psfrag{D}[cr]{\txt{$\phantom{x}$\\Symbol Ticks}}
\psfrag{I}[bl]{\txt{\Dgreen $\phantom{x}$Classification \\ \Dgreen By Annihilator}}
 \subfloat[]{\includegraphics[width=3.1in]{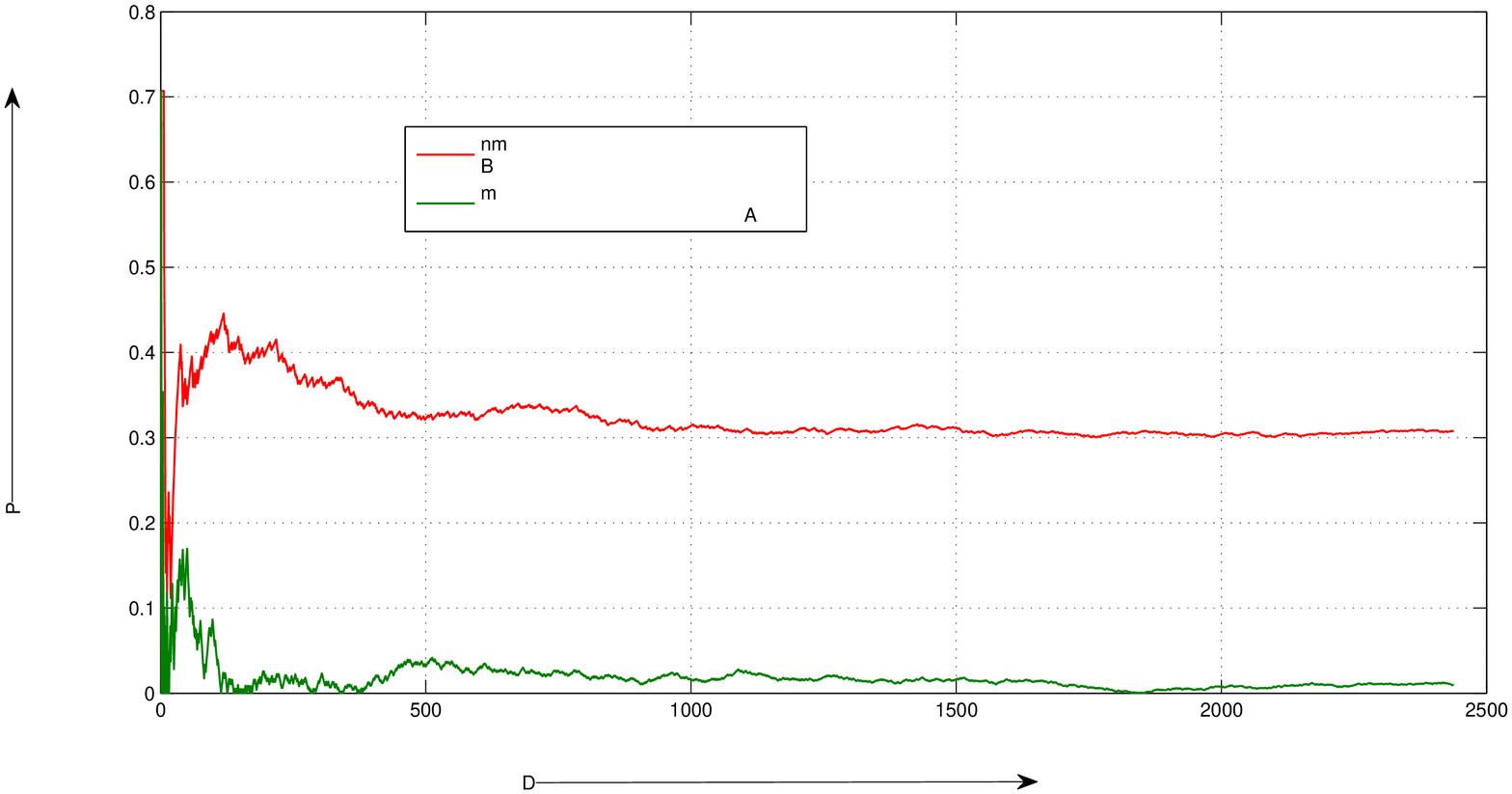}}
%
%
%
\caption{(a) Simulation result for model M2: Note that direct compression converges at $3500$ symbols, while the annihilator converges at just over $1000$ symbols. A few intermediate models
identified by the direct compression algorithm is also shown. (b) Simulation result for model S1: Note that direct compression has failed to converge at $7000$ symbols; (c) Comparison of distance of annihilated strings from symbolic white noise for  correct and incorrect annihilators, $i.e.$, where we have a pattern match and where we do not.
}\label{figsim1}
\end{figure*}

The algorithm used for direct compression is a modified version of CSSR~\cite{SS04,SSC02}. We compare PFSA models using the metric $\theta$ proposed in \cite{CR08}, which is capable of computing
distances between PFSA models with different underlying graphs (with identical alphabets). \textit{Note that while the output of the direct compression algorithm is compared against the original model, the annihilator output is compared against symbolic white noise.}

In the two simulation runs reported, we generate data from the models and compare the string lengths required by direct compression versus classification via annihilation. The annihilators
were constructed from the knowledge of the particular model used in the simulation using the formulation presented in Section~\ref{subsecann}. Note, that the annihilation technique is not meant for
identification of an unknown pattern ($i.e.$ pattern identification), but detecting if the sensed symbol string is actually being generated by a known library pattern ($i.e.$ pattern classification). Figure~\ref{figsim1}(a) illustrates the results for M2, and we note that the annihilator is significantly faster.
 The principal advantage of using  annihilators is better illustrated for S1, where, for reasons explained above, the direct compression
algorithm has a hard time, and has failed to correctly identify S1 even after $7000$ symbols (convergence was observed at around $10000$ symbols). The annihilator identifies S1 at just over $1000$ symbols as shown in Figure~\ref{figsim1}(b). Finally, Figure~\ref{figsim1}(c) compares the response of a annihilator which does not correspond to the process generating the observed symbols, with one that does. Note, that in both cases, the responses are very stable; with the incorrect annihilator converging to $0.3$ and the correct one to (very nearly) $0$, which reflects a match. In general, evaluation of such a pattern match involves using  a specified detection threshold, the implications of which are discussed in the next section.

Implication of Proposition~\ref{propbounds} is illustrated explicitly in Figure~\ref{figschemesensa} (a snapshot of the annihilation process in the above described simulation runs), where we note that the annihilation process essentially erases symbols selectively in the incoming data stream, and hence yields a significantly shorter observed sequence. Although, we now have the relatively easier task of identifying if this annihilated sequence is indeed white; but even such an identification cannot be effectively done with too few symbols. Proposition~\ref{propbounds} guarantees that the length shortening cannot offset this advantage in practical scenarios, where we are most likely to have more states than the total number of symbols in the alphabet.
\vspace{0pt}

\section{Intuitive Interpretation \& Potential Applications}\label{secdiscuss}

\begin{figure*}[t]
\centering
\subfloat[]{\begin{minipage}{\textwidth}\parashade[.9]{sharpcorners}{\hspace{-8pt}
\begin{tabular}{ll}
{\footnotesize  \itshape \bf \scshape Original}: & \hspace{-10pt}\scriptsize  \BRed  110000000011110000001100000110011000000001101100001100001100001111000011001100110011000000110001100010000000001110001111\\
{\footnotesize  \itshape \bf \scshape Annhltd}: & \hspace{-10pt}\scriptsize  \color{DarkGreen} 1{\color{white}0}0{\color{white}0}{\color{white}0}{\color{white}0}{\color{white}0}{\color{white}0}{\color{white}0}{\color{white}0}1{\color{white}0}110{\color{white}0}0{\color{white}0}{\color{white}0}{\color{white}0}1{\color{white}0}0{\color{white}0}{\color{white}0}{\color{white}0}{\color{white}0}1{\color{white}0}{\color{white}0}{\color{white}0}1{\color{white}0}{\color{white}0}{\color{white}0}{\color{white}0}{\color{white}0}000{\color{white}0}1{\color{white}0}{\color{white}0}1{\color{white}0}0{\color{white}0}{\color{white}0}0110{\color{white}0}0{\color{white}0}1{\color{white}0}{\color{white}0}{\color{white}0}{\color{white}0}{\color{white}0}{\color{white}0}{\color{white}0}11{\color{white}0}{\color{white}0}0{\color{white}0}1{\color{white}0}0{\color{white}0}1{\color{white}0}001{\color{white}0}{\color{white}0}{\color{white}0}1{\color{white}0}{\color{white}0}{\color{white}0}{\color{white}0}{\color{white}0}{\color{white}0}{\color{white}0}1{\color{white}0}{\color{white}0}{\color{white}0}0{\color{white}0}{\color{white}0}{\color{white}0}{\color{white}0}{\color{white}0}{\color{white}0}0{\color{white}0}{\color{white}0}0{\color{white}0}{\color{white}0}{\color{white}0}{\color{white}0}{\color{white}0}{\color{white}0}{\color{white}0}1{\color{white}0}{\color{white}0}{\color{white}0}1{\color{white}0}11\\
{\footnotesize \itshape \bf \scshape  Observd}: & \hspace{-10pt}\scriptsize \bf \color{DodgerBlue} 10111001011000110011001110101001110001111 $\phantom{X}$\colorbox{white}{\color{Red4}\bf \footnotesize \sffamily $\longleftarrow$ Length Shortening Of Observed Annihilated Stream }
\end{tabular}}\end{minipage}\label{figschemesensa}
}\\
\subfloat[]{\psfrag{P}[lc]{ \txt{\Dgreen Pattern \\\Dgreen Library }}
\psfrag{C}[cl]{\footnotesize \color{Red4} \bf \sffamily \txt{Compute Match \\ (Via\\ Structural\\ Comparison)}}
\psfrag{L}[cb]{\scriptsize \bf \Mblue \txt{Pattern Library \\ (Constructed Via Compression of Training Data)}}
\psfrag{Cx}[cr]{\footnotesize \Nblue \bf \sffamily \txt{Symbolization \color{Green4} (Easy) \\ \& Compression Algorithm \color{Red4}(Expensive)}}
\psfrag{M}[bc]{\footnotesize \bf \sffamily \itshape\textshade{sharpcorners}{\txt{Computed \\ Model}}}
\psfrag{C2}[ct]{\footnotesize \color{Red4} \txt{Expensive}}
\psfrag{S}[cc]{\bf \color{Green4} \txt{Sensor Signal}}
\psfrag{N}[lc]{\footnotesize \txt{ Comparison}}
 \includegraphics[width=6in]{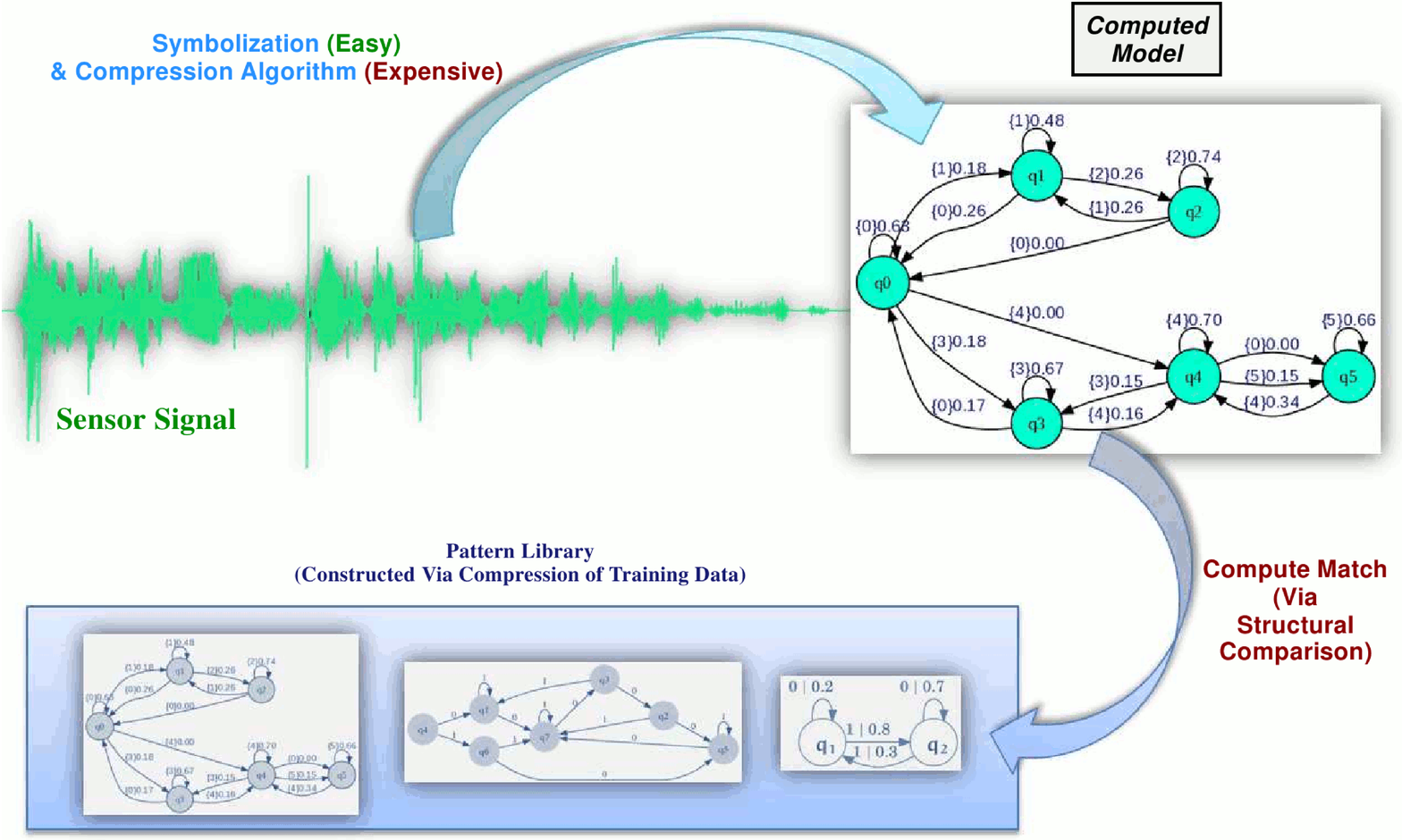}\label{figschemesensb}}
\caption{(a) Snapshot of the annihilation process (b) Pattern classification scheme in symbolized sensory data streams}\label{figschemesens}
\end{figure*}
An important intuitive insight on why the annihilators are able to classify the streams faster can be given as follows: by avoiding direct compression of the observed symbol sequence, we are essentially solving a classification problem, which is, in general, easier compared to a full blown identication problem, involving discovery of new  patterns. Direct compression is capable of telling us not only if there is a match, but also yields the new PFSA model  of the observed sequence when there is no match with the existing templates. Annihilation only indicates the matching template if there is one, and indicates a "no match" otherwise. Thus, the increased efficiency is not surprising. This is particularly useful for templates that have no synchronizing strings (such as the model $S1$), where for direct compression, one needs to distinguish between the states using long observation sequences, that disambiguate possible future evolutions based on the small deviations in the observed probability distributions on the future strings. If the spectral gap of the corresponding Markov chain is small, the sequences required can turn out to be unacceptably long (since smaller the spectral gap, longer is the mixing time). This is what we see manifested in Figure~\ref{figsimTb}, where  direct compression has a hard time. For the annihilation, such complexities are absent: the spectral gap does play a role in the degree of shortening of the annihilated sequence (See Proposition~\ref{propbounds}), but one always looks for symbolic white noise at the annihilator output, irrespective of the complexity of the template.

The idea of pattern classification via controlled information erasure may seem somewhat counter-intuitive at the first reading. However, the key notion exploited here has a clear  analogue in  communication theory, particularly in the theory of matched filters~\cite{N63}. 
A matched filter is a theoretical construct (and not the name of a specific  filter family) which processes a received signal to minimize the effect of noise, $i.e.$ maximizes the signal to noise ratio (SNR), and simultaneously minimizes the probability of bit error rate (BER).
It can be shown that, under the assumption of additive white Gaussian  noise (AWGN) in the communication channels,  an optimum filter for receiver-end demodulation exists, and is
a function only of the transmitted pulse shape. Because of this direct relationship to the
 transmitted pulse, it is called a {\itshape matched filter}. 
The derivation of a classical matched filter is essentially based on a direct application of  Schwartz
inequality~\cite{R88}, and  leads to a very simple and remarkable
conclusion:
\begin{quote} \itshape
For AWGN channels, the signal to noise ratio is maximized when the impulse response of that filter is
exactly a reversed and time delayed copy of the transmitted signal.
\end{quote}
Since the bit error rate experienced by a signal during
demodulation is a function of the signal to noise ratio~\cite{Pr00}, a matched filter which maximizes
SNR will automatically provide the lowest possible BER. 
The analogy of semantic annihilation with matched filters is compelling: instead of using a time-reversed copy of the signal template, we are using the symbol stream generated by an inverse probabilistic automata. Just as a matched filter functions by convolving the signal with its reversed and delayed copy, the annihilator carries out symbol-wise comparisons between the given symbol stream, and the state-specific ones generated by the inverted template; erasing symbols that do not match. The fact that we can carry out this procedure in a  deterministic fashion should not be surprising: the convolution in the case of matched filters is generally carried out using Fourier transforms (FFT), which is also a rather straightforward deterministic operation. In the latter case, the filtered signal must still  be recognized, but this decision-making task is now significantly easier due to the filtration-enhanced SNR. In our case, the annihilator does not output an enhanced signal, but reduces it to white noise if the correct template is used. However, the task of recognizing symbolic white noise is significantly easier compared to recognizing the template pattern directly; thus reinforcing the analogy. The recognition of symbolic white noise does involve the use of a detection threshold, since in practical scenarios, we do not expect the signal and the template to match exactly, given finite-length observation sequences. Thus, when the distance between at least one of the PFSA models computed from the annihilator output falls within a pre-specified distance to the white noise model (in the sense of the PFSA metric $\theta$~\cite{CR08}), we conclude a positive match. Using arbitrarily small thresholds may require long data streams, and most likely will result in negative matches due to small noise-mediated  mismatch between the streams.

The key application that the authors have in mind is pattern classification in symbolized (or quantized) sensory data streams. This particular approach of pattern detection in sensory data has been shown to be significantly more efficient to classical continuous domain techniques, exhibiting remarkable insensitivity to spurious 
noise and exogenous disturbances;  primarily due to the quantization-mediated coarse-graining, and 
 as a consequence of repeated recurrences of paths in the graph of the finite 
state machine with relatively few states and a large number of sample points in the (fast scale) 
time series data~\cite{R04}. Recent applications of such PFSA-based pattern classification has been effectively applied to anomaly detection problems in complex electro-mechanical machines~\cite{CRR05}, and tracking targets via large-scale multi-modal urban sensor networks~\cite{CWPR10}.
The basic philosophy is illustrated in Figure~\ref{figschemesensb}. Continuous valued data from sensor(s) is quantized via an appropriately chosen partitioning scheme~\cite{RR06} to yield a symbolic sequence over a pre-specified alphabet (depending on the coarseness of the chosen partition). In the absence of annihilators, one is then required to algorithmically compress a sufficiently long symbolic sequence to extract the underlying causal generative model in the form of  a probabilistic finite automata. The classifier is provided with a template library consisting of  PFSA models that encode the pertinent patterns of interest. Once the observed sequence is compressed to a PFSA, this can then be compared against the individual library elements to compute a possible match. The compression algorithms, however, are often expensive; particularly if the underlying PFSA is not a subshift of finite type~\cite{LM95}. Annihilation offers a significantly simple solution, which skips the compression step altogether. The observed stream can be symbol-wise annihilated using the inverted templates in the library, requiring less data, and significantly simpler implementations.

A second promising application  is the design of PFSA-based novel modulation-demodulation schemes for communication over noisy channels.
In this paper, we considered the special case 
where the symbol stream generated by a PFSA $G$ is annihilated by the inverse model $-G$. However, in general,  one can apply similar ideas to encode a stream from PFSA $G$ using an encoding PFSA $G_e$ as $G\pfsum G_e$, and demodulate by "adding" the inverse stream $G\pfsum G_e \pfsum (-G_e)=G$. Such avenues will be explored in future, where careful choice of the encoding PFSA may lead to greater resilience to noise corruption, or even to unauthorized message access.

\section{Summary, Conclusions \& Future Work}\label{secsumm}
We defined an additive abelian group for probability measures on symbolic strings, which induces an abelian group on a slightly restricted set of PFSA.
The defined PFSA sum is then  used to formulate semantic annihilators, which identify pre-specified patterns of interest via perfect removal of all inter-symbol correlations from observed strings, turning them to symbolic white noise. This approach of classification via annihilation is shown to be advantageous, with theoretical guarantees, for a large class PFSA models. The results are supported by simulation experiments.

Future work will extend the formulation to models where not all symbols satisfy the condition that the generation  probabilities are strictly non-zero from each model state.
The effect of noise corruption on observed strings need to be investigated, with particular emphasis on the comparative effect of noisy observations on direct compression and semantic annihilation. Furthermore, implementation in actual experimental scenarios will further validate the proposed classification technique.
}
%
\bibliographystyle{IEEEtran}
\bibliography{BibLib1}
\end{document}